\documentclass[a4paper,11pt]{article}
\pdfoutput=1 

\usepackage{jcappub} 
\usepackage{latexsym,amssymb,amsmath}
\usepackage{enumerate}
\usepackage{csquotes}
\usepackage{multirow}
\usepackage[dvipsnames]{xcolor}

\title{Neural network reconstruction of the dense matter equation of state from neutron star observables}

\author[a,b,c]{Shriya Soma}
\author[a,c]{Lingxiao Wang}
\author[d,e]{Shuzhe Shi}
\author[a,b,f]{Horst St\"ocker}
\author[a,1]{Kai Zhou\note{Corresponding author.}}

\affiliation[a]{Frankfurt Institute for Advanced Studies (FIAS), D-60438 Frankfurt am Main, Germany}
\affiliation[b]{Institut f\"ur Theoretische Physik, Goethe Universit\"at, D-60438 Frankfurt am Main, Germany}
\affiliation[c]{Xidian-FIAS International Joint Research Center, D-60438 Frankfurt am Main, Germany}
\affiliation[d]{Department of Physics, McGill University, Montreal, Quebec H3A 2T8, Canada.}
\affiliation[e]{Center for Nuclear Theory, Department of Physics and Astronomy, Stony Brook University, Stony Brook, New York, 11794, USA.}
\affiliation[f]{GSI Helmholtzzentrum f\"ur Schwerionenforschung GmbH, D-64291 Darmstadt, Germany}

\emailAdd{soma@fias.uni-frankfurt.de}
\emailAdd{lwang@fias.uni-frankfurt.de}
\emailAdd{shuzhe.shi@stonybrook.edu}
\emailAdd{stoecker@fias.uni-frankfurt.de}
\emailAdd{zhou@fias.uni-frankfurt.de}

\abstract{The Equation of State (EoS) of strongly interacting cold and hot ultra-dense QCD matter remains a major challenge in the field of nuclear astrophysics. With the advancements in measurements of neutron star masses, radii, and tidal deformabilities, from electromagnetic and gravitational wave observations, neutron stars play an important role in constraining the ultra-dense QCD matter EoS. In this work, we present a novel method that exploits deep learning techniques to reconstruct the neutron star EoS from mass-radius (M-R) observations. We employ neural networks (NNs) to represent the EoS in a model-independent way, within the range of $\sim$1-7 times the nuclear saturation density. The unsupervised Automatic Differentiation (AD) framework is implemented to optimize the EoS, so as to yield through TOV equations, an M-R curve that best fits the observations.
We demonstrate that this method works by rebuilding the EoS on mock data, i.e., mass-radius pairs derived from a randomly generated polytropic EoS. The reconstructed EoS fits the mock data with reasonable accuracy, using just 11 mock M-R pairs observations, close to the current number of actual observations.}


\begin{document}
\maketitle
\flushbottom

\section{Introduction}
Neutron stars (NSs) harbour extreme conditions that are unattainable in terrestrial laboratories and are therefore instrumental in constraining the cold dense matter equation of state (EoS). The past decade has witnessed great progress in the research of these objects. NS observables like mass, radius, moment of inertia and tidal deformability are crucial in probing the EoS of strongly interacting dense matter. The mass measurements of pulsars have reached a high level of precision with observations via post-Keplerian parameters, e.g. Shapiro delay. Some of the most recent and accurate mass measurements include PSR J0348+0432, which has a mass of 2.01$\pm$0.04~M$_{\odot}$~\cite{antoniadis:2013massive}, PSR J0740+6620, with a mass of 2.08$\pm$0.07~M$_{\odot}$~\cite{Fonseca_2021}, and J1810+1714, with a mass of 2.13$\pm$0.04~M$_{\odot}$~\cite{Romani_2021}. The detection of gravitational waves from binary neutron star mergers (BNSMs) by the LIGO-Virgo collaboration has also made it possible to extract yet another parameter, the tidal deformability ($\Lambda$), of excited neutron stars for the very first time. The analysis of the first BNSM collision event, GW170817, which was later accompanied by an electromagnetic counterpart, has led to an estimated value of, $197\leq \Lambda\leq 720$~\cite{coughlin:2018constraints, ligoPhysRevX.9.011001}. Furthermore, towards the end of 2019, NICER (Neutron Star Interior Composition Explorer) announced its first results on radius measurements of a pulsar, namely PSR J0030+0451~\cite{riley:2019nicer,miller:2019psr}. With an expected substantial increase in observational data, it is possible to establish further constraints on the dense matter EoS. 
Experiments at the GSI Helmholtzzentrum f\"ur Schwerionenforschung and the future Facility for Antiproton and Ion Research (FAIR) in Europe pave the way to compress baryonic matter to high densities and pressures through nucleus-nucleus collisions. The temperatures in these collisions are however high~\cite{FRIESE2006377}. We therefore resort to observational data from cold, long lived NSs, particularly masses and radii, to further this study.

It has been demonstrated in previous studies that there exists a one-to-one mapping from the mass-radius (M-R) relationship of neutron stars to the EoS~\cite{lindblom:1992determining}. This is realized by inverting the Tolman--Oppenheimer--Volkoff (TOV) equations~\cite{PhysRev.55.364, PhysRev.55.374}, which are numerically solved to obtain the structural properties of NSs from a given EoS. With sufficiently many NS observations, spread across the M-R space, one can, in this way, try to extrapolate the NS EoS to the data from finite nuclei at low densities, and to perturbative QCD calculations at asymptotically high densities.
Theoretical models which can generate possible EoSs that attempt to describe the dense matter inside NSs incorporate different physical assumptions for the strongly interacting dense matter. Non-relativistic and relativistic model EoSs may include only purely nucleonic degrees of freedom, hybrid (hadrons and quarks) models, models with hyperons, models with kaon condensates and pure quark models~\cite{baym1971ApJ.170.299B, Steiner_2013, antonPhysRevC.101.034904, Alford_2005, Blaschke:2013CQ, BALBERG1997435, Banik_2014, Malik_2021, Lattimer_2001}. These EoSs can be used to predict the global properties of NSs, which can then be confronted with observational data. This allows to constrain NS parameters like maximum mass, radius, and tidal deformability, which are all dependent on the EoS~\cite{Rezzolla_2018, PhysRevD.100.023015, eliasPhysRevLett.120.261103, Soma_2020, ligoPhysRevX.9.011001}. This method is however strongly model-dependent. It does not systematically translate existing NS observations to dense matter properties. Alternatively, EoSs are parameterized with piecewise polytropes~\cite{readJPhysRevD.79.124032, raithel:2016neutron}, spectral representations~\cite{LindblomPhysRevD.82.103011}, or Gaussian processes~\cite{Han:2021kjx, legred2022implicit}. 

Previous attempts to reconstruct the EoS from M-R relations in a hopefully model-independent way, for example, rely on the well-known Bayesian inference. This approach utilizes a certain parameterization for the EoSs and applies particular choices for their priors~\cite{Steiner_2010, raithel:2017neutron, Traversi_2020}. 
One possible way to exploit directly the non-linear mapping between an M-R curve and its EoS, is the machine learning inference in the sense of supervised learning. Recent studies based on this strategy include works by Fujimoto et al.~\cite{fujimoto:2018methodology, fukushimaPhysRevD.101.054016, fujimoto:2021extensive}, Morawski et al.~\cite{morawski:2020neurala}, Ferreira and Provid{\^{e}}ncia~\cite{Ferreira_2021}, and Krastev~\cite{Krastev:2021reh}. Fujimoto et al. developed a feed forward neural network which outputs the EoS in its speed of sound representation, when supplied with M-R observations. Morawski et al. used the encoder-decoder structure of the Auto-Encoder to reconstruct the EoS, without using any parametric representation. Ferreira and Provid{\^{e}}ncia used Support Vector Machines to regress the EoS in terms of nuclear matter parameters. Krastev showed that, from observational NS data, a trained feed forward neural network is capable of extracting the density dependence of the nuclear symmetry energy and therefore the EoS.
In the present study, a model-independent reconstruction of the dense matter EoS is accomplished with statistical inference from M-R observations in an unbiased manner. The method introduced here utilizes deep neural networks (DNNs) in the Automatic Differentiation (AD) framework to reconstruct the NS EoS. This is the first implementation of an unsupervised learning algorithm for the EoS reconstruction. Additionally, the design of our inverse problem necessitates a functional form of the TOV equations that is easily differentiable. For this purpose, we train a DNN to solve the TOV equations, essentially modeling a \texttt{TOV-Solver Network}. The framework is tested on mock data generated using piece-wise polytropes for the EoS. 

The layout of the article is as follows: Section~\ref{dl-methods} outlines the paradigm of Automatic Differentiation with DNNs in the reconstruction procedure of the NS EoS. Section~\ref{tov-solver-section} provides a description of the \texttt{TOV-Solver Network} and, accordingly, the data preparation process. Section~\ref{eos-network} details the specifics of the \texttt{EoS Network} and its optimization. The results are discussed in Section~\ref{results} and the conclusions are drawn in Section~\ref{conclusions}.

\begin{figure*}
\begin{center}
\includegraphics[width=0.75\textwidth]{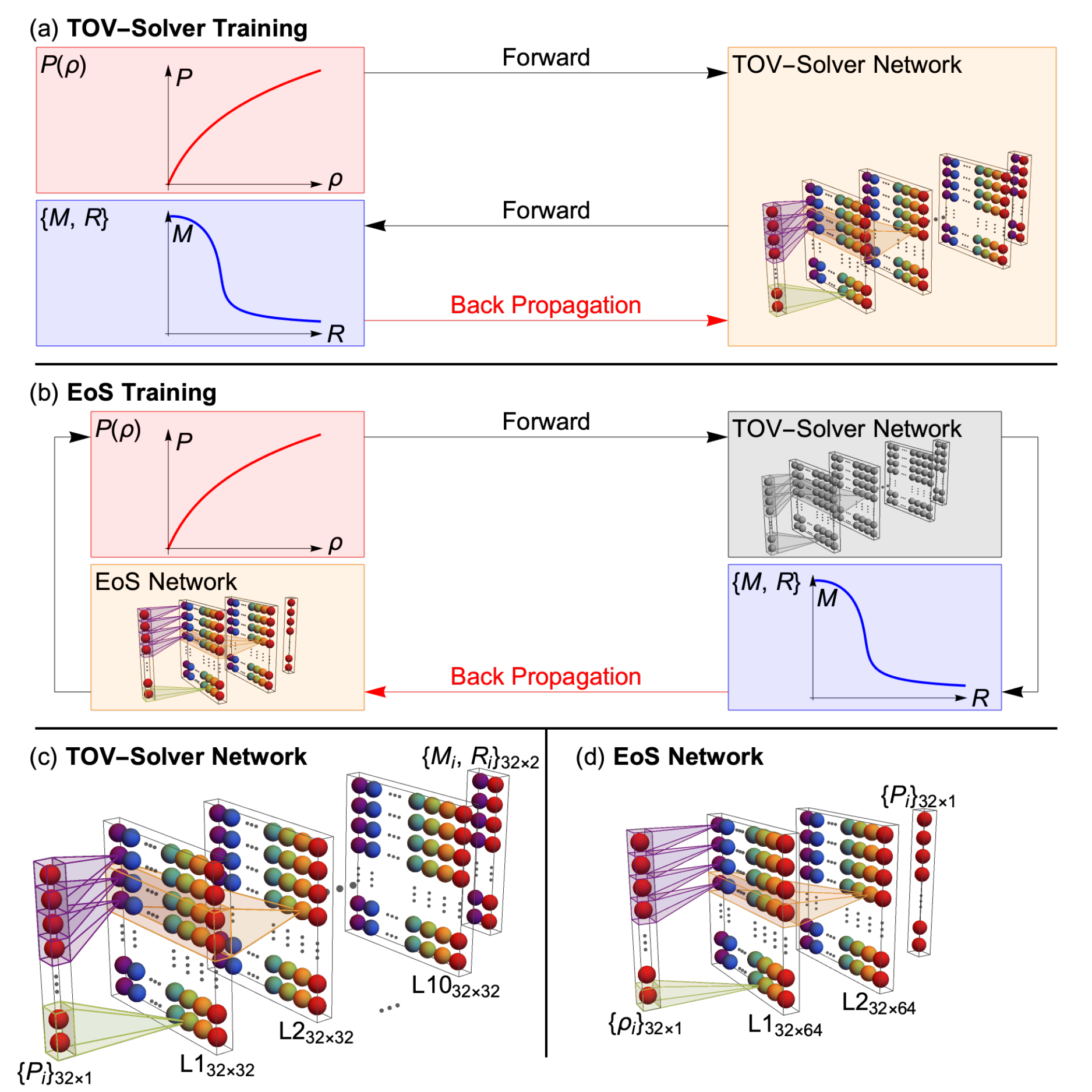}
\caption{A flow chart of our methods, with TOV-Solver training (a) and EoS training (b). Note that in (b) the \texttt{TOV-Solver Network} is well-trained and frozen. The structure of the \texttt{TOV-Solver Network} and \texttt{EoS Network} are depicted in (c) and (d) respectively, in which, the colored spheres are nodes of the network, and the colors label the indices across the width in each layer. The shadow lines link the nodes between layers, and represent a 1D convolutional kernel operation with trainable parameters.}\label{completescheme}
\end{center}
\end{figure*}

\section{Reconstructing the EoS via Automatic Differentiation} \label{dl-methods}
Due to their abilities to capture complex nonlinear correlations in data, deep learning techniques have been proven useful in solving a number of physical problems, e.g. determining the parton distribution function~\cite{Forte_2002, Collaboration_2007}, reconstructing the spectral function~\cite{Kades:2019wtd, Zhou:2021bvw,Chen:2021giw}, identifying phase transitions~\cite{Carrasquilla_2017,Pang:2016vdc,Wang:2020hji, Du:2019civ, Wang:2020tgb, Jiang:2021gsw}, assisting lattice field theory calculations~\cite{Boyda:2020hsi, Zhou:2018ill, Kanwar:2020xzo, Albergo:2019eim}, evaluating centrality distributions for heavy ion collisions~\cite{OmanaKuttan:2020brq, Thaprasop:2020mzp, Li:2020qqn}, parameter estimation under detector effects~\cite{Andreassen:2020gtw, Kuttan:2020mqj}, and speeding up hydrodynamic simulations~\cite{Huang:2018fzn}.
Earlier works that incorporated DL methods have shown that DNNs can potentially surpass traditional methods in solving inverse design problems~\cite{doi:10.1126/sciadv.aar4206, shi2021heavy, wang2021reconstructing}. 
In the course of this work, we explore the DNNs' capacities to invert the TOV equations, in order to reconstruct the NS EoS from a limited set of M-R observations, through an unsupervised Automatic Differentiation framework. 

Figure~\ref{completescheme} summarizes the flow chart of our approach: Part(a) of figure~\ref{completescheme} illustrates a primary DNN that is trained, supervised, to emulate the numerical methods used to solve the TOV equations, i.e. mapping a given EoS to its corresponding M-R curve, hence the name \texttt{TOV-Solver Network}~\footnote{The universal approximation theorem ensures that DNNs can approximate any kind of continuous function with nonlinear activation functions~\cite{goodfellow:2016deep}.}. Once trained, the network model is saved as a mapping from x to $\mathbf{z}$, where $\mathbf{z}=f(x)$. Here, $x= P_i(\rho_i)$ is the pressure at fixed mass densities, which is the input to the \texttt{TOV-Solver Network}. The output, $\mathbf{z}=(M_i,R_i)$, represents the mass-radius pairs. The network model is composed of a series of differentiable modules which includes both linear transformations and nonlinear activation functions. In Figure~\ref{completescheme}(b), we inherit the well-trained \texttt{TOV-Solver Network} from the previous step with all its differentiable parameters frozen in subsequent procedures. A secondary DNN, the \texttt{EoS Network}, is introduced to represent the EoS in an unbiased and flexible manner, i.e. $P_{\theta}(\mathbf{\rho})$, where $\{\theta\}$ is a set containing all parameters of the \texttt{EoS Network}. By coupling to the well-trained \texttt{TOV-Solver Network}, the \texttt{EoS Network} is optimized in an unsupervised manner to fit the combined output to the M-R observations. Hence, given the number of M-R observations, $N_{\text{obs}}$, the loss function for training is the standard $\chi^2$ between the observations and the predictions of the pipeline ``\texttt{EoS Network} $\to$ \texttt{TOV-Solver Network}'', and is given by
\begin{equation}
\chi^2 = \sum_{i=1}^{N_{\text{obs}}} \frac{(M_{i} - M_{\text{obs},i})^2}{\Delta M_i^2}
+    \frac{(R_{i} - R_{\text{obs},i})^2}{\Delta R_i^2}. \label{eq:chi2}
\end{equation}
Here, ($M_{i},R_i$) is the predicted output of the observation  ($M_{\text{obs},i},R_{\text{obs},i}$), which has an uncertainty ($\Delta M_{i},\Delta R_i$). Given a frozen \texttt{TOV-Solver Network}, we can derive the gradients of its parameters as
\begin{equation}
\frac{\delta\chi^2}{\delta \theta} = \frac{\delta\chi^2}{\delta \mathbf{z}} \frac{\delta \mathbf{z}}{\delta P_{\theta}}  \frac{\delta P_{\theta}}{\delta \theta}.
\end{equation}
Here, the last two terms can be directly computed using a back-propagation algorithm~\cite{goodfellow:2016deep} within the AD framework for the two coupled DNNs. In other words, we are essentially fine-tuning the parameters of the \texttt{EoS Network} to obtain the desired M-R curve, after going through the frozen \texttt{TOV-Solver Network}, with guidance from a limited set of observational data. The architectures of the \texttt{TOV-Solver Network} and the \texttt{EoS Network} are shown in Figure~\ref{completescheme} as part~(c) and (d), respectively. We utilized the Python Library Keras~\cite{chollet2015keras}, which is built on the Tensorflow platform~\cite{tensorflow2015-whitepaper}, to setup the network models and to perform the AD calculations for optimizing the NS EoS.

\section{TOV-Solver Network} \label{tov-solver-section}
This section describes the \texttt{TOV-Solver Network} devised for the present work. The network is built on a supervised learning scheme with the purpose of efficiently solving the TOV equations, i.e. to obtain the M-R curve from an arbitrary EoS. We therefore set the EoSs as the input to the network and train the network to output the corresponding M-R sequences. The training and validation of the \texttt{TOV-Solver Network} demands sufficient data. In the following subsections, we provide information on the data-generation process implemented in our study. Subsection~\ref{poly-eos} is devoted to the preparation of polytropic EoSs, subsection~\ref{tov-eq} to the TOV equations and their solutions, and subsection~\ref{tov-net-res} to the training of the \texttt{TOV-Solver Network} and subsequent investigation of its performance.

\subsection{Piecewise Polytropic Equations of State} \label{poly-eos}
 
A multitude of equations of state (EoSs) is generated spanning wide ranges in the pressure-density $(P-\rho)$ plane. This is achieved by parameterizing the EoSs in terms of piecewise polytropes. The low-density part of each EoS ($\rho < \rho_{\text{sat}}$, where $\rho_{\text{sat}}\sim2\times10^{14}~\text{g~cm}^{-3}$ is the nuclear saturation density) is assumed to conform to a conventional nuclear EoS. Nevertheless, in order to introduce robustness in the training dataset, the low-density regime of the EoSs in the present study is assumed to comply with one of SLy~\cite{douchin:2001unified}, PS~\cite{PANDHARIPANDE1975507} or DD2~\cite{PhysRevC.81.015803} EoSs. For the high-density region ($\rho > \rho_{\text{sat}}$), we adopt the density segmentation scheme from Ref~\cite{raithel:2016neutron}, i.e. any EoS can be reasonably well parameterized with five polytropic segments. The high-density region of the EoSs is chosen within the range [$\rho_{\text{sat}}$, 7.4$\rho_{\text{sat}}$]. This upper limit follows from the evidence that the pressure at this density, $P(\rho = 7.4\rho_{\text{sat}})$, affects the maximum mass of the neutron star~\cite{ozel:2009prd}. The density is uniformly spaced on a logarithmic scale. The five segments are separated at densities (1.0, 1.4, 2.2, 3.3, 4.9, 7.4) $\rho_{\text{sat}}$, as in Ref.~\cite{raithel:2016neutron}. The pressure in the $i^{\text{th}}$ segment is given as a function of the density $\rho$, 
\begin{equation}\label{eq1}
P = K_i \rho^{\Gamma_i} \hspace{0.3cm} \text{for i = [1,5],}
\end{equation}
where,
\begin{enumerate}[(i)]
  \item ~$\rho \in [\rho_{i-1}, \rho_{i}]$,  with $\rho_{i-1}$ and $\rho_i$ being the minimum and maximum densities of segment i. 
  \item $K_i = P_{i-1}/\rho_{i-1}^{\Gamma_i}$.
  \item Following Ref.~\cite{raithel:2016neutron}, the adiabatic index, $\Gamma_i$ is assigned random values in the range [1,~$\min$\{5,~$\Gamma_{\text{luminal}}$\}), where $\Gamma_{\text{luminal}}$ sets the limit for the causal condition, i.e. the speed of sound, $c_s$, does not exceed the speed of light, c. Hence, $c_s = \sqrt{dP/d\epsilon} < 1$, or
  \begin{equation}
      \Gamma \equiv \Gamma_{\text{ luminal}} \quad\text{when}\quad \frac{dP}{d\epsilon} = 1.
  \end{equation}
\end{enumerate}
\begin{figure}[b]
\begin{center}
\includegraphics[width=0.55\columnwidth]{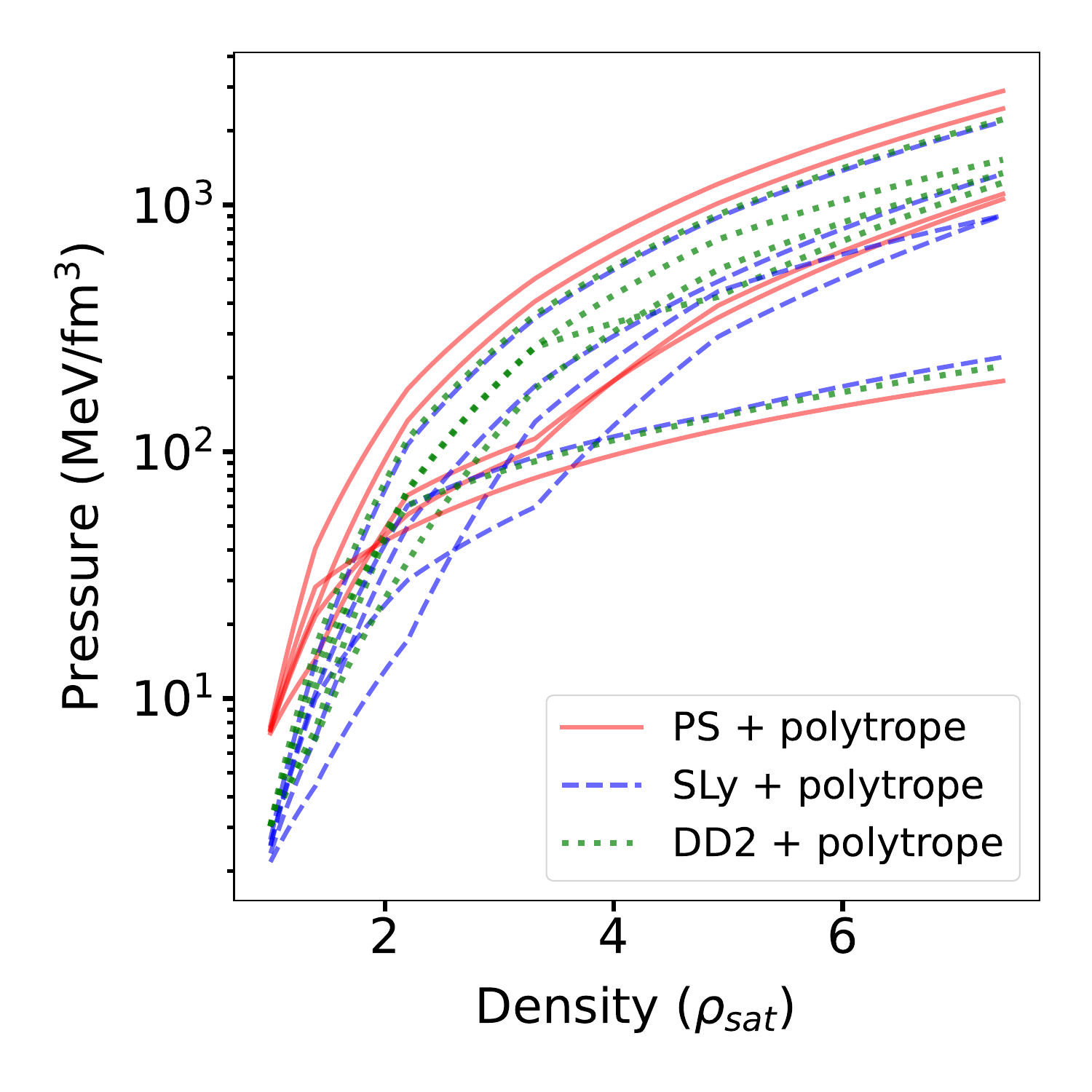}
\caption{A few piece-wise polytropic EoSs that were generated using PS (red), SLy (blue) or DD2 (green) EoSs for the low-density region. A major portion of the pressure range is spanned by the EoSs.}\label{eos_poly}
\end{center}
\end{figure}
Here, $\epsilon$ is the energy density, given by
\begin{equation}
    \epsilon = \left(\frac{\epsilon(\rho_{i-1})}{\rho_{i-1}} - \frac{P_{i-1}}{(\Gamma_i - 1)\rho_{i-1}}\right)\rho + \frac{K_i}{\Gamma_i - 1}\rho^{\Gamma_i}\,,
\end{equation}
for $\Gamma \neq 1$, and 
\begin{equation}
    \epsilon = \frac{\epsilon(\rho_{i-1})}{\rho_{i-1}}\rho + K_i \ln\left(\frac{1}{\rho_{i-1}}\right)\rho - K_i \ln\left(\frac{1}{\rho}\right)\rho\,,
\end{equation}
for $\Gamma=1$ (see Ref.~\cite{raithel:2016neutron} or~\cite{ozel:2009prd} for details).

With this prescription, three sets of 100,000 polytropic EoSs for each category, SLy, PS and DD2, are generated. A few of the generated EoSs are depicted in Figure~\ref{eos_poly}.

\subsection{TOV Equations: From EoS to Stellar Structure} \label{tov-eq}
\begin{figure}[!hbtp]
\begin{center}
\includegraphics[width=7.5cm]{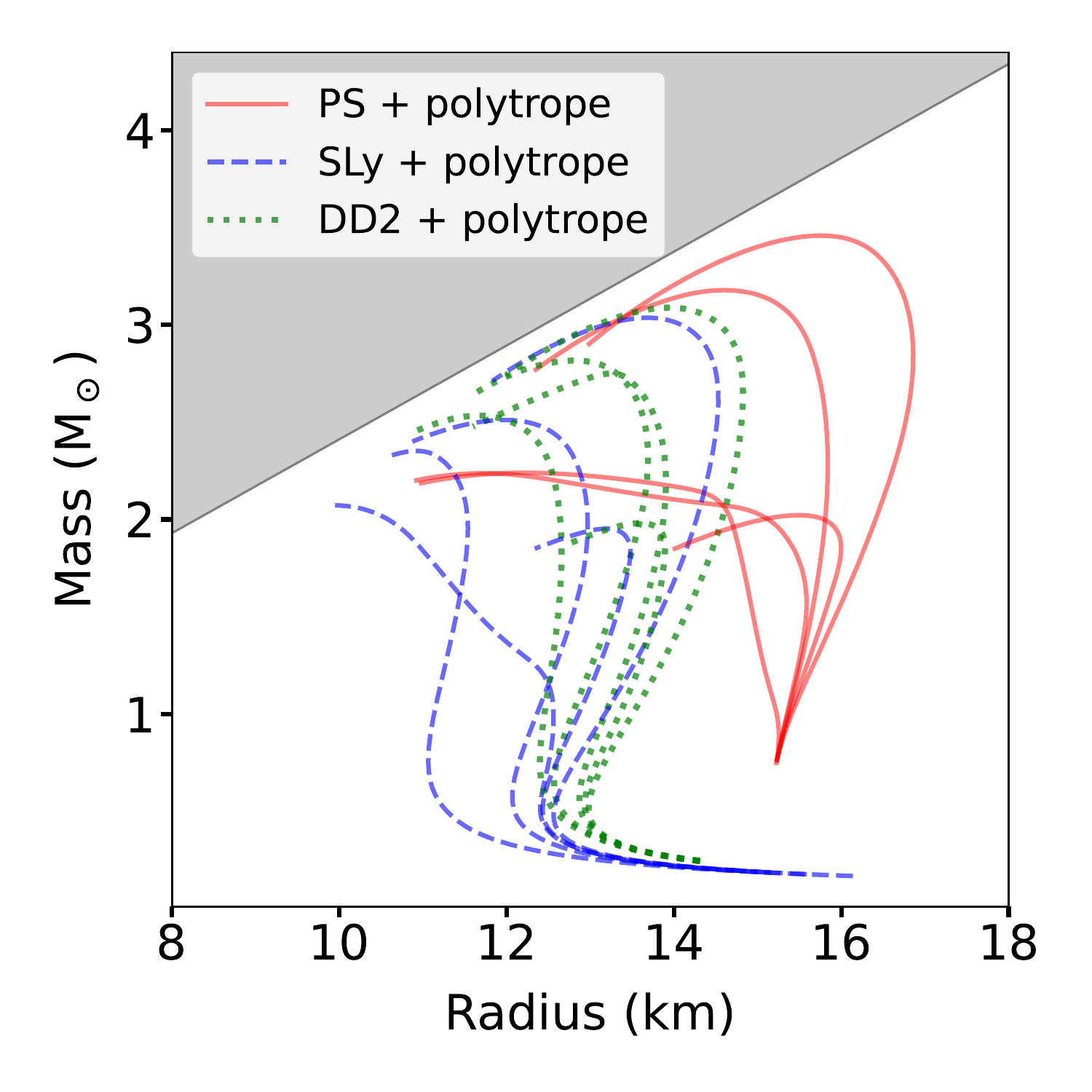}
\caption{The Mass-Radius curves correspond to the EoSs plotted in Figure~\ref{eos_poly}, obtained by solving the TOV equations. The gray line denotes the causal limit.}\label{mr_poly}
\end{center}
\end{figure}

The Tolman--Oppenheimer--Volkoff (TOV) equations are derived from the Einstein equations for a spherically symmetric star in hydrostatic equilibrium~\cite{PhysRev.55.364, PhysRev.55.374}. They are given as
\begin{equation}
    -\frac{dP}{dr} = \frac{\big{[}\epsilon(r)+P(r)\big{]}\big{[}m(r)+4\pi r^3 P(r)\big{]}}{r[r-2m(r)]}\,,
\end{equation}
and
\begin{equation}
\frac{dm(r)}{dr} = 4\pi r^2 \epsilon (r).
\end{equation}
Here, $r$ is the radial distance from the centre of the star, and $m(r)$ is the mass enclosed within the radial distance, $r$. In order to determine the observables, mass ($M$) and radius ($R$) of the star, the TOV equations are integrated radially outwards from the centre. The initial conditions taken at the centre of the star are, $r=0, ~m(r=0)=0, ~\epsilon(r=0)\neq0, ~P(r=0)=P_c$, where $P_c$ is the central pressure, obtained from the EoS, usually given as a table. The radius, $R$, of the star is defined by the vanishing pressure condition at the surface ($P(r=R)=0$), and the mass enclosed in $R$ is the total mass of the star, i.e. $M=m(R)$. Thus, the mass-radius (M-R) sequences are calculated for all the EoSs generated in subsection~\ref{poly-eos}, a few of which are depicted in Figure~\ref{mr_poly} (corresponding to the EoSs in Figure~\ref{eos_poly}). 
From this set of data, we exclude all the EoSs (and their corresponding M-R sequences) which fail to accommodate a neutron star of mass 1.9$M_{\odot}$. This preference of a conservative limit follows from the observations~\cite{Demorest2010, antoniadis:2013massive, Fonseca_2021}, and leaves us with 94,462 EoSs corresponding to PS; 61,273 EoSs corresponding to SLy; and 72,834 EoSs corresponding to DD2 for training and validating the \texttt{TOV-Solver Network}. \\

\subsection{Training the TOV-Solver Network} \label{tov-net-res}
The simulated data from subsections~\ref{poly-eos} and~\ref{tov-eq} are used to train the \texttt{TOV-Solver Network}. The deep learning model, WaveNet, is adopted to perform this emulator task. WaveNet is a generative neural network model with autoregressive properties. It mimics the concept of autoregression which is exercised in solving the TOV equations (see subsection~\ref{tov-eq}). Therefore, it is used here to resemble the conventional numerical methods. 
The EoSs which are input to the \texttt{TOV-Solver Network} are represented in a discretized format, $P_i \equiv P(\rho_i)$. The contribution to the EoSs from the low density region ($\rho<\rho_{sat}$) are omitted in the network as these values have been established in our work to follow one of the three conventional nuclear EoSs: PS, SLy or DD2. Thus, we designate the input layer as $P_i \equiv P(\rho_{sat}\leq\rho_i \leq7.4\rho_{sat})$, a single channel, i.e, an array of pressure, with shape $(N_{\rho},1)$, where $N_{\rho}$ is the number of discrete density values. 
The output however has a shape $(N_{\rho},2)$, i.e, two channels for the mass $M$, and radius, $R$, respectively. Two different resolutions for the EoS representation are used here, namely $N_\rho$ = 128 or 32. It is realized in subsequent assessments that a coarse resolution suffices to accomplish the desired results. The model uses logarithmic values of pressure as the input. Both the input and the output arrays are normalized to constitute elements that lie within the range (0,1). Additional information on the network design and training is supplied in Appendix~\ref{appendixA}
\footnote{A few additional DL models, namely, the fully connected neural network (FCN), Convolutional neural network (CNN), and Long Short Term Memory (LSTM), were also evaluated based on their execution of this task. Appendix~\ref{appendixA} entails a description of their structures and an analysis of their performances.}.
\begin{figure}[!hbtp]
\centering
\includegraphics[width=0.5\columnwidth]{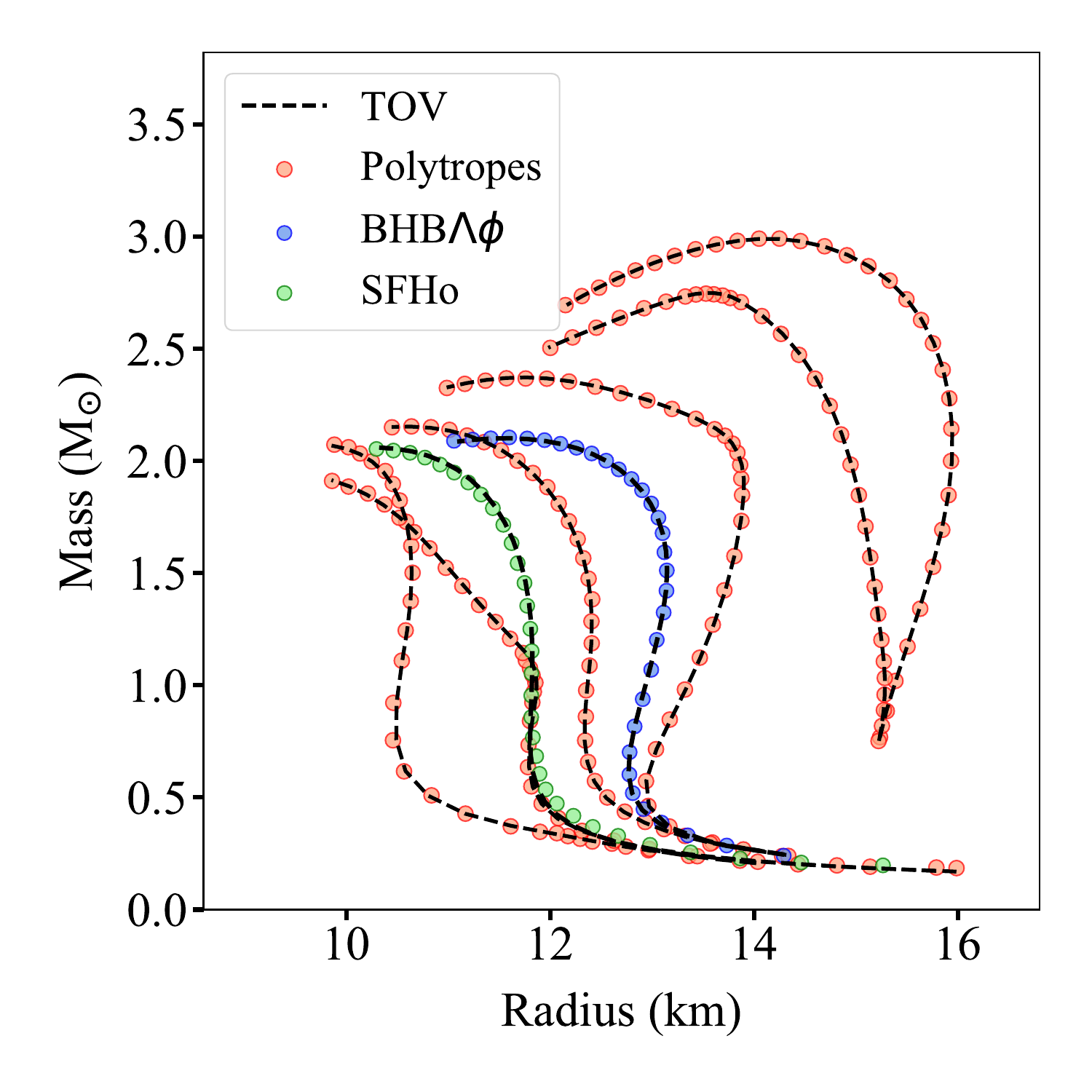}
\caption{The \texttt{TOV-Solver Network} predictions using the WaveNet model for several unseen test EoSs are shown against the ground truth values of the M-R curves. The dashed black line is output from TOV equations and the markers denote the network predictions. The blue and green dotted M-R curves represent the network predictions for BHB$\Lambda\phi$ and SFHo EoSs respectively. }\label{NN_pred}
\end{figure}
Figure~\ref{NN_pred} shows that the trained \texttt{TOV-Solver Network} is successful in capturing the mapping of randomly given EoSs to the corresponding M-R curves. Note that the trained \texttt{TOV-Solver Network} also works well on SFHo and BHB$\Lambda\phi$ EoSs as shown in Figure~\ref{NN_pred}. The M-R curves are successfully reproduced by the DL models to a very high precision, with the coefficient of determination $\mathcal{R}^2\sim 99.9\%$. This illustrates that DNNs have the capacity to replace conventional numerical methods used for solving the TOV equations. In comparison to numerical calculations like the Euler or Runge-Kutta methods, the network emulator used here, is (i) superior in computational efficiency, ($\sim10^6$ {\it sec} faster), and
(ii) easily differentiable. This is critical for applying back-propagation in the AD framework, deployed in the following section for statistical inference of the EoS reconstruction. The trained \texttt{TOV-Solver Network} model and weights are saved. 
\section{EoS Network} \label{eos-network}

The \texttt{EoS Network} is introduced into the pipeline at this stage (see Figure~\ref{completescheme}(b)). It is modeled with density, $\rho$, as input and the corresponding pressure, $P_{\theta}(\rho)$, as output. The EoS output ($P_{\theta}(\rho)$) from the \texttt{EoS Network} is further input to the well-trained \texttt{TOV-Solver Network}, thus linking the two networks. The \texttt{TOV-Solver Network} parameters are set as non-trainable weights prior to the EoS optimization process. The trainable weights of the \texttt{EoS Network} are then optimized in this pipeline to fit the predicted M-R output to mock observational data, thereby inverting the TOV equations. Compared to earlier studies which captured the inverse mapping from M-R observations to the EoS~\cite{fujimoto:2018methodology, fukushimaPhysRevD.101.054016, fujimoto:2021extensive,morawski:2020neurala, Ferreira_2021,Krastev:2021reh}, the proposed method belongs to the unsupervised learning paradigm, which can be cast as a generalized Bayesian inference, with augmentations in the following aspects: (i) the EoS is represented in an unbiased manner as a DNN, thus the parameters to optimize are the network weights and biases; (ii) the traditional numerical methods to solve the TOV equations are replaced by a well-trained \texttt{TOV-Solver network}, thus simplifying and speeding up the following AD process; and (iii) the optimization uses a gradient-descent based approach within the AD framework as depicted in Figure~\ref{completescheme}(b).
The network architecture and the optimization procedure are given below.

The input density to our pipeline is a linearly spaced 1D array of length $N_{\rho}$= 32, normalized to lie within $(0,0.1)$. The trained WaveNet model of the same resolution is deployed as the \texttt{TOV-Solver Network}~\footnote{All the trained DL models were tested in the pipeline. Optimum results were obtained with the WaveNet and LSTM}. 
The \texttt{EOS Network} architecture utilizes three 1D convolutional layers (two hidden layers, and one output layer) as depicted in Figure~\ref{completescheme}(d). The hidden layers have 64 feature maps each. A kernel of size 1$\times$1 is used, with the kernel weights initialized from a {\it{He}} normal distribution~\cite{he2015delving}. The $L_2$ regularizer, $\lambda=10^{-8}$, is exercised on the weights, and the {\it same} padding is applied to all layers. The Sigmoid activation function is used on the output layer, and the ELU activation function on the hidden layers. By using kernel size 1$\times$1, we ensure that each input element shares the same parameters in the above architecture. Therefore, the induced relationship between the $i^{\text{th}}$ input density neuron, $\rho_i$, and corresponding output pressure, $P_i$, is equivalent to a fully connected DNN representation with two hidden layers consisting of $64$ neurons each. Thus, the \texttt{EoS Network} uses a total of 4353 parameters for defining $P_{\theta}(\rho)$.
We further specify the weights of each layer in the \texttt{EoS Network} to be non-negative. This preserves the order of the input layer up to the output layer, $P(\rho)$. Thus, we establish monotonicity in the represented function $P(\rho)$, a condition that is required by any physical EoSs. This method also ensures that the reconstructed EoS is a well correlated function. 
Contrary to the general use of large datasets in training a neural network, the optimization procedure in our approach requires just one M-R sequence. Each training epoch therefore has a batch size 1, i.e, in the ideal situation, we are required to only optimize the unique dense matter EoS to fit M-R observations. However, we adapt to the realistic scenario where all observational data comes with sizeable uncertainties (further details in Section~\ref{results}).

Assuming $N_{\text{obs}}$ is the number of reliable M-R observations, the loss function for the training is given by Eq.~(\ref{eq:chi2}). It is defined as the distance between the observations and the M-R curve as predicted from the network pipeline above, i.e. the likelihood of observations given an EoS and its corresponding M-R curve from the \texttt{TOV-Solver Network}. However, the observational data is not spread uniformly across the M-R space, and the measurement uncertainties are likely to cause discontinuities in the M-R curve. Furthermore, the data is limited and the central density, ($\rho_{ci}$), to which an uncertain observation corresponds to, is unknown. In other words, the $i^{\text{th}}$ M-R observation, ($M_i,R_i$), does not necessarily correspond to the $i^{\text{th}}$ central density, $\rho_{ci}$, in the input layer~\footnote{In such circumstances, a finer resolution ($N_{\rho}=128$) might prove useful.}. To evaluate the loss function optimally, we adopt the `closest approach' method as implemented in Ref.~\cite{raithel:2017neutron}. Thus, within each iteration during the training, we evaluate the loss as 
\begin{equation}
    \chi^2 = \sum_{i=1}^{N_{\text{obs}}} \frac{(M(\rho_{ci}) - M_{\text{obs},i})^2}{\Delta M_i^2}
+    \frac{(R(\rho_{ci}) - R_{\text{obs},i})^2}{\Delta R_i^2}\,.
\label{eos_loss}
\end{equation}
Here, $\rho_{ci}$ for every $i^{\text{th}}$ observation is updated as,
\begin{equation}
    \rho_{ci} =\arg \min_{\rho_c} \frac{(M(\rho_{c}) - M_{\text{obs},i})^2}{\Delta M_i^2}
+    \frac{(R(\rho_{c}) - R_{\text{obs},i})^2}{\Delta R_i^2}\,.
\label{indexupdate}
\end{equation}
Eq.~(\ref{indexupdate}) is therefore used to determine the central densities of ($M_{\text{obs}},R_{\text{obs}}$) that lead to the least distance between the M-R curve obtained from the \texttt{TOV-Solver Network} and the M-R observations. 
The Adam optimizer~\cite{DBLP:journals/corr/KingmaB14} was utilized here, with varying learning rates ($\alpha$'s) for different stages of the training process where smaller learning rates are usually used at later stages to stabilize the training. The learning progress of the network is moderated by decreasing the loss function $\chi^2$. The learning rate for one such reconstruction in the present study, is scheduled as follows: 1000 epochs ($\alpha$ = 0.001), 1000 epochs ($\alpha$ = 0.005), 1500 epochs ($\alpha$ = 0.003), 1600 epochs ($\alpha$ = 0.0001) and, finally, 2000 epochs ($\alpha$ = 0.00003), in the order specified. The reason for the small changes in the learning rates is that they have minor effects on the final results. Using this method, the EoS is reconstructed with an uncertainty which is proportional to the statistical uncertainty of the observations. In certain cases where the causal condition is not fulfilled, the DNN fails to produce a realistic EoS. This is also seen from the lack of convergence of the loss during the optimization process. This issue is resolved by applying the causal condition to reject those EoSs.

\section{Results and Discussion} \label{results}
\begin{figure}[htbp!]
\begin{center}
\includegraphics[width=0.95\columnwidth]{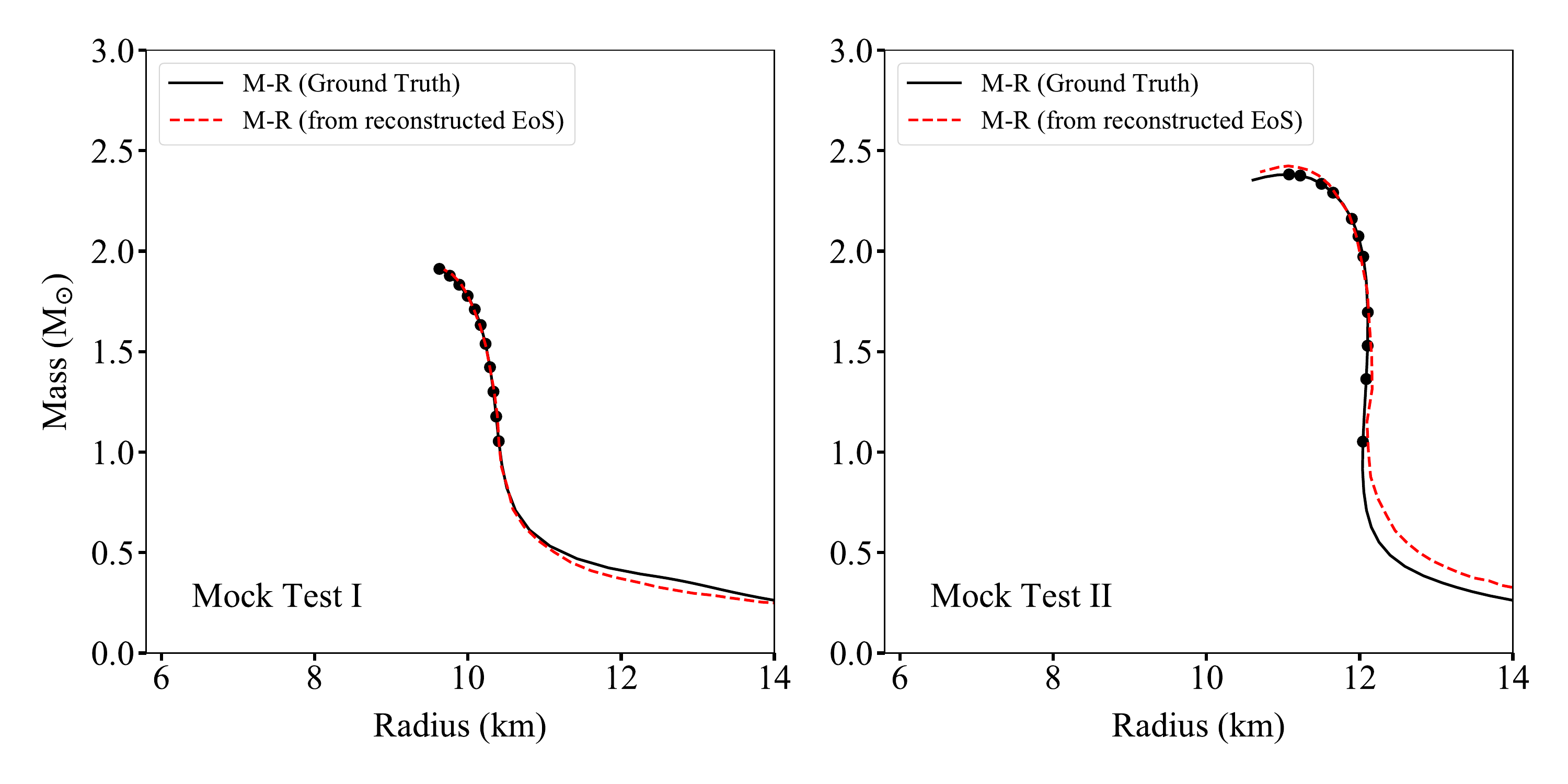}
\caption{The 11 M-R mock data (black points) located in the region $M>1M_{\odot}$ along the ground truth M-R curve (black solid line). A reasonable agreement of the M-R curve from the reconstructed EoS (red dashed line) with the ground truth curve is depicted in the mass region $M>1M_{\odot}$. The left and right panels respectively represent the Mock Test I and Mock Test II.}\label{idealMR}
\end{center}
\end{figure}

\begin{figure}[htbp!]
\begin{center}
\includegraphics[width=0.95\columnwidth]{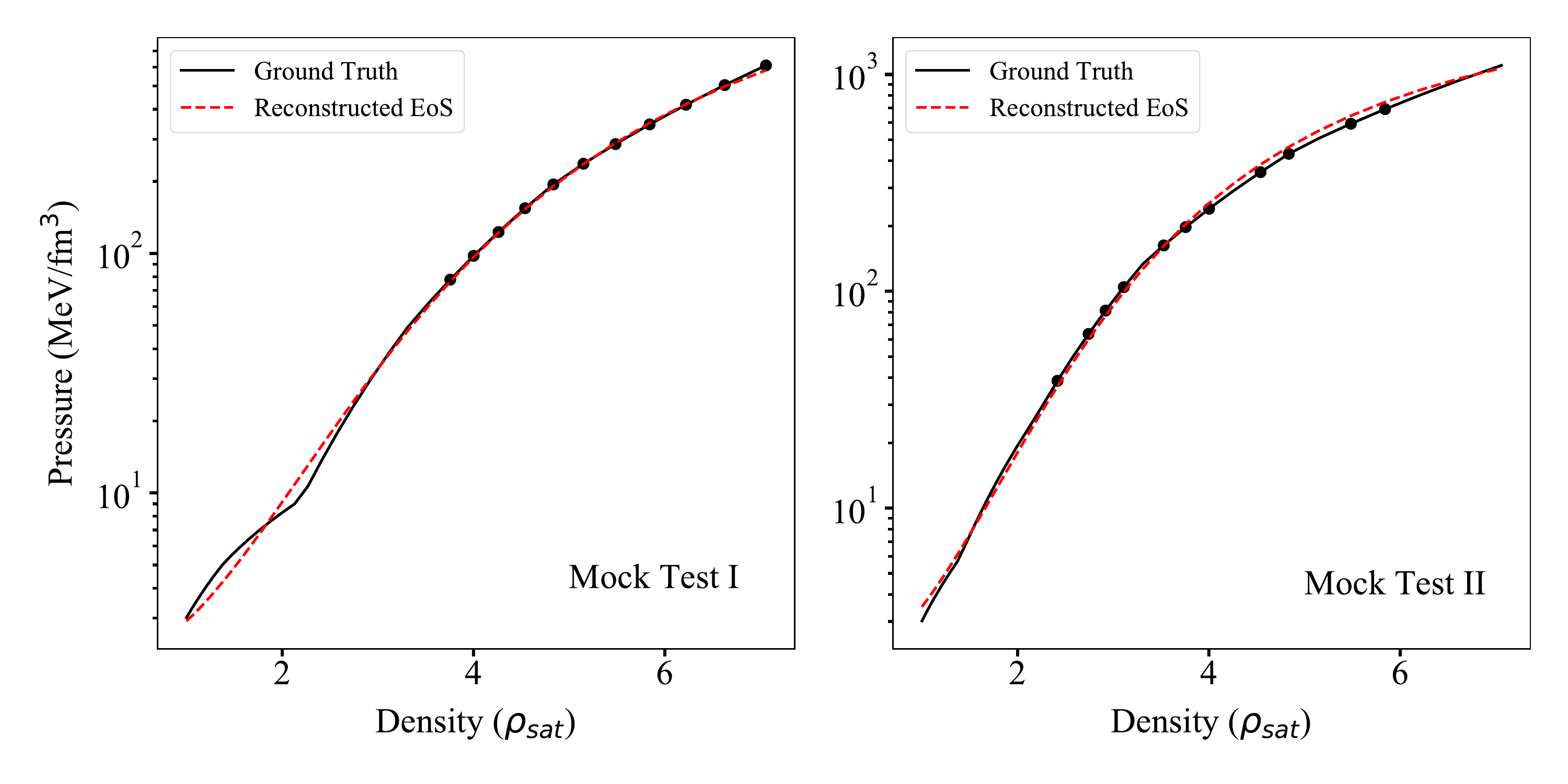}
\caption{The comparison of the EoS reconstructed by our method (dashed red line) and the ground truth (solid black line). The black points shown, when taken as central densities for the NSs, correspond to the mock M-R ``observations'' in Figure~\ref{idealMR}. The left and right panels respectively represent the Mock Test I and Mock Test II.}\label{idealEoS}
\end{center}
\end{figure}

The reconstruction method proposed here is examined by closure tests that are first performed in an ideal case of mock data, i.e, synthetic M-R mock data without systematic and statistical uncertainties. We use two sets of such mock data from M-R curves which correspond to two randomly chosen EoSs as depicted in Figure~\ref{idealEoS}. The black solid lines in the figures are the targets for the reconstruction attempt. The absence of low mass NSs ($<1M_{\odot}$) in nature compels us to use mock data with masses above $1M_{\odot}$. Figure~\ref{idealMR} depicts two sets of such mock observations each containing 11 M-R points, which are chosen from the region $M>1M_{\odot}$ along the M-R curve.
These 11 highlighted black points lie on the M-R curve (ground truth). 
With our proposed method, the EoSs from these two sets of mock M-R data are reconstructed. Figure~\ref{idealEoS} displays a comparison of the EoS reconstructed by the DNN in this work (dashed red line) with the ground truth (solid black line). Evidently, the reconstruction of the EoSs is fairly successful in the range $M>1M_{\odot}$ for the ideal scenarios. The reconstruction of the EoSs by the DNN is remarkably close to the ground truth in the high-density region (i.e., corresponding to the high mass region in M-R curve). The deviation of the reconstructed EoSs from the ground truth in the low-density regime is attributed to the lack of mock data below 1$M_{\odot}$. This is also demonstrated in Figure~\ref{idealEoS}, where the 11 black solid points on the EoSs correspond to the 11 mock data points on the M-R curves in Figure~\ref{idealMR}. The red dashed curves in Figure~\ref{idealMR} are the M-R curves corresponding to the reconstructed EoSs as predicted by the \texttt{TOV-Solver Network}. The two M-R curves agree with each other reasonably well. For a sanity check, one can obtain the M-R curve of the reconstructed EoS by directly solving the TOV equations and compare the respective results. 

\begin{figure}[htbp!]
\begin{center}
\includegraphics[width=0.55\columnwidth]{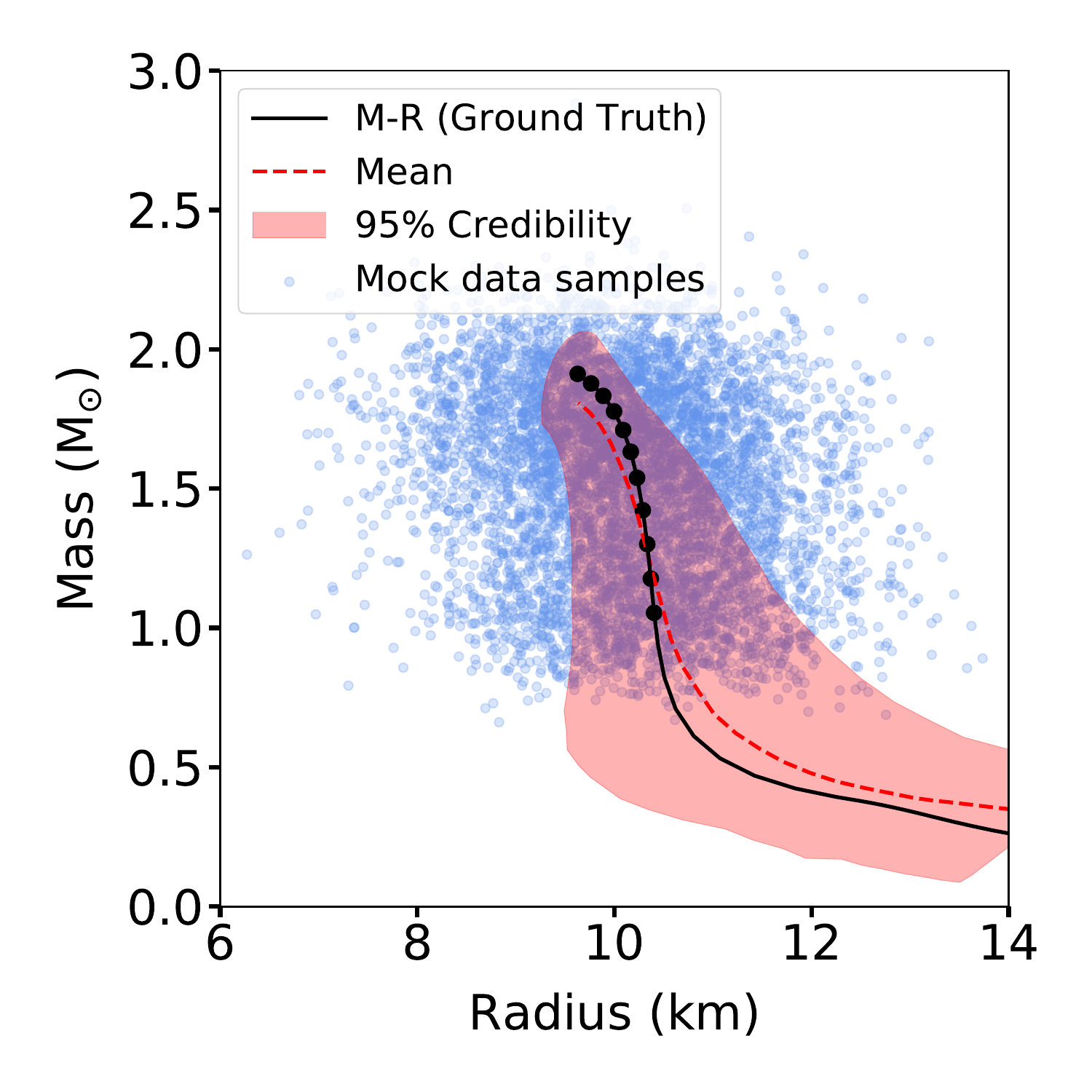}
\caption{The blue point cloud represents the M-R mock data ensembles w.r.t mock test I, sampled with $10\%$ uncertainties from the ground truth M-R curve (black solid line). The $95\%$ confidence level of the M-R band from the reconstructed EoSs is depicted as the orange shaded region, and the mean as a red dashed line.}\label{rec_eos_mr}
\end{center}
\end{figure}

\begin{figure}[htbp!]
\begin{center}
\includegraphics[width=0.55\columnwidth]{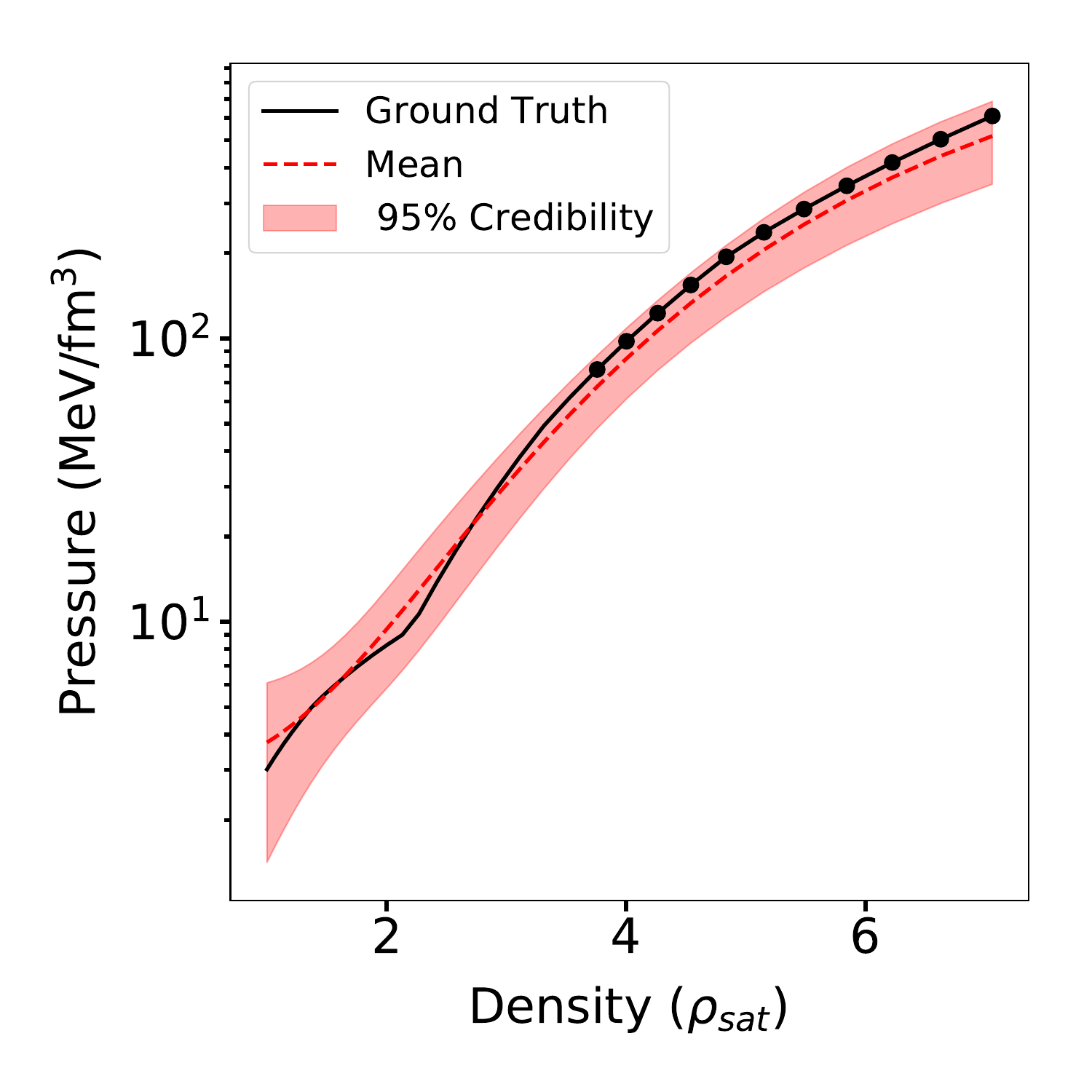}
\caption{Comparison of the ground-truth (black solid line) and the reconstructed EoS (red dashed line) with uncertainty (orange band) by using M-R mock data from Figure~\ref{rec_eos_mr}.}\label{rec_eos}
\end{center}
\end{figure}

We proceed to the realistic situation where the observations inevitably involve uncertainties. The proposed method is tested on the same mock M-R data as in Figure~\ref{idealMR}, albeit with uncertainties included. To incorporate statistical noise, for each of the two mock tests, multiple M-R curves are sampled from a normal distribution around the true M-R values, $\mathcal{N}(M_i;R_i;\sigma_{M_i};\sigma_{R_i})$. A 10\% relative standard deviation is assigned to each mass and radius, i.e $\Delta M_i=0.1M_i$ and $\Delta R_i=0.1R_i$. Therefore, 500 mock M-R data samples are drawn from a normal distribution with $\sigma_M = 0.1M$, $\sigma_R = 0.1R$. In Figure~\ref{rec_eos_mr}, the true M-R curves are represented by the black solid line and the highlighted circles mark the 11 M-R mock data, which are the mean values of the respective normal distributions. The blue point-cloud in Figure~\ref{rec_eos_mr} depicts all 500 sampled sets of the mock data points resulting from the true M-R curve, where each sampled set contains just 11 M-R pairs, all above $\sim1M_{\odot}$, to be consistent with the current real observations. The optimization procedure is repeated for the 500 samples and the results are shown in Figure~\ref{rec_eos}. For each sampled set of mock data, the proposed method reconstructs exactly one EoS in the sense of maximum a posteriori probability (MAP). In this closure test, the uncertainty of the reconstruction is evaluated by fitting the reconstructed EoSs to a uni-variate normal distribution, $\mathcal{N}(\mu_i,\sigma_i)$. This is straightforward and a $2\sigma$, 95\% confidence level of the reconstructed, model-independent EoS is shown in Figure~\ref{rec_eos} as the shaded region. 
The dashed red curve in the shaded region is the mean value of the reconstructed EoSs.
The \texttt{TOV-Solver Network} predicts the M-R curves corresponding to each of the reconstructed EoSs. The M-R predictions of the \texttt{TOV-Solver Network} from the reconstructed EoSs are then subjected to a normal distribution fit. In this case, all M-R points with the same central density $\rho_{c,i}$ (i.e. the ensemble of each element from the output sequence) are fitted with a separate bi-variate Gaussian distribution, given by $p((M_i,R_i)|{P_j}) = \mathcal{N}(\mathbf{\mu_i},\mathbf{\sigma_i})$. The 95\% confidence level is determined from the $2.44\sigma_i$ interval for 2D distributions. This uncertainty band of the M-R curve is depicted by the orange band in Figure~\ref{rec_eos_mr}. The resulting mean M-R curve from the bi-variate distribution is represented by the dashed red line. When compared to the uncertainty of the mock M-R data samples (blue point clouds), it is evident that the width of the M-R curves obtained from the reconstructed EoSs is considerably smaller. This illustrates the potential of the present novel EoS reconstruction method - it works despite the large uncertainties in the mock M-R data. For a few sample M-R curves, the NN fails to produce a realistic EoS. The causal condition is applied to filter out these EoSs which fail to converge. In certain cases, a failure can also be ascribed to the large uncertainties of individual M-R pairs which leads to outliers of the M-R curve. Moreover, the points on the M-R curve are assumed to fall in the ascending order of their mean mass. However, accounting for large uncertainties while sampling from a Gaussian distribution, may result in a disordering of the M-R pairs.


\begin{figure}[htbp!]
\begin{center}
\includegraphics[width=0.95\columnwidth]{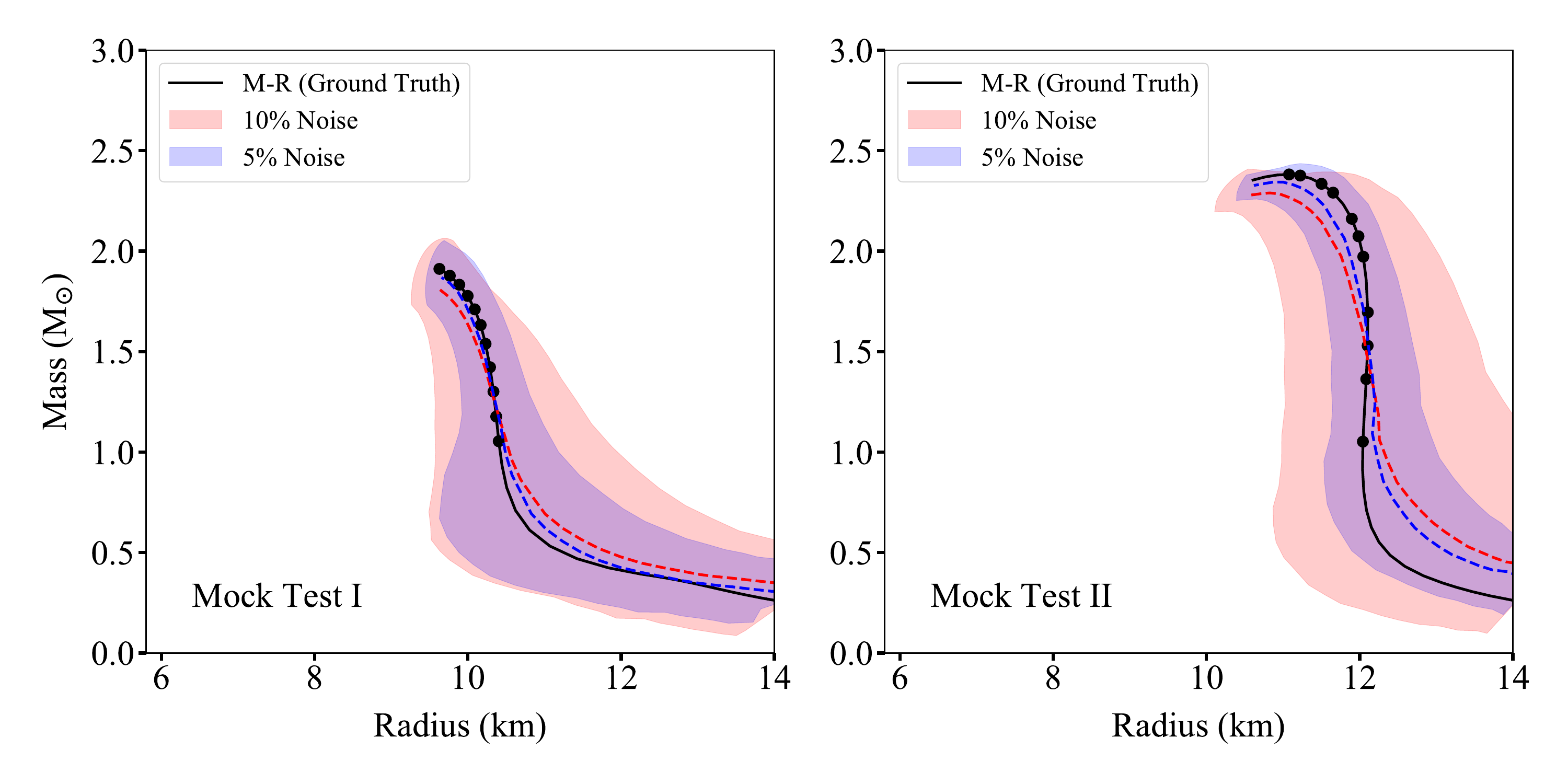}
\caption{The ground truth M-R curve (black solid line) and the M-R relationship from reconstructed EoSs with mean (red dashed line for $10\%$ noise level, blue for $5\%$) and uncertainty (orange band for $10\%$ noise level and blue for $5\%$) depicted. The left and right panels respectively represent the Mock Test I and Mock Test II.}\label{rec_eos_mr2}
\end{center}
\end{figure}

\begin{figure}[htbp!]
\begin{center}
\includegraphics[width=0.95\columnwidth]{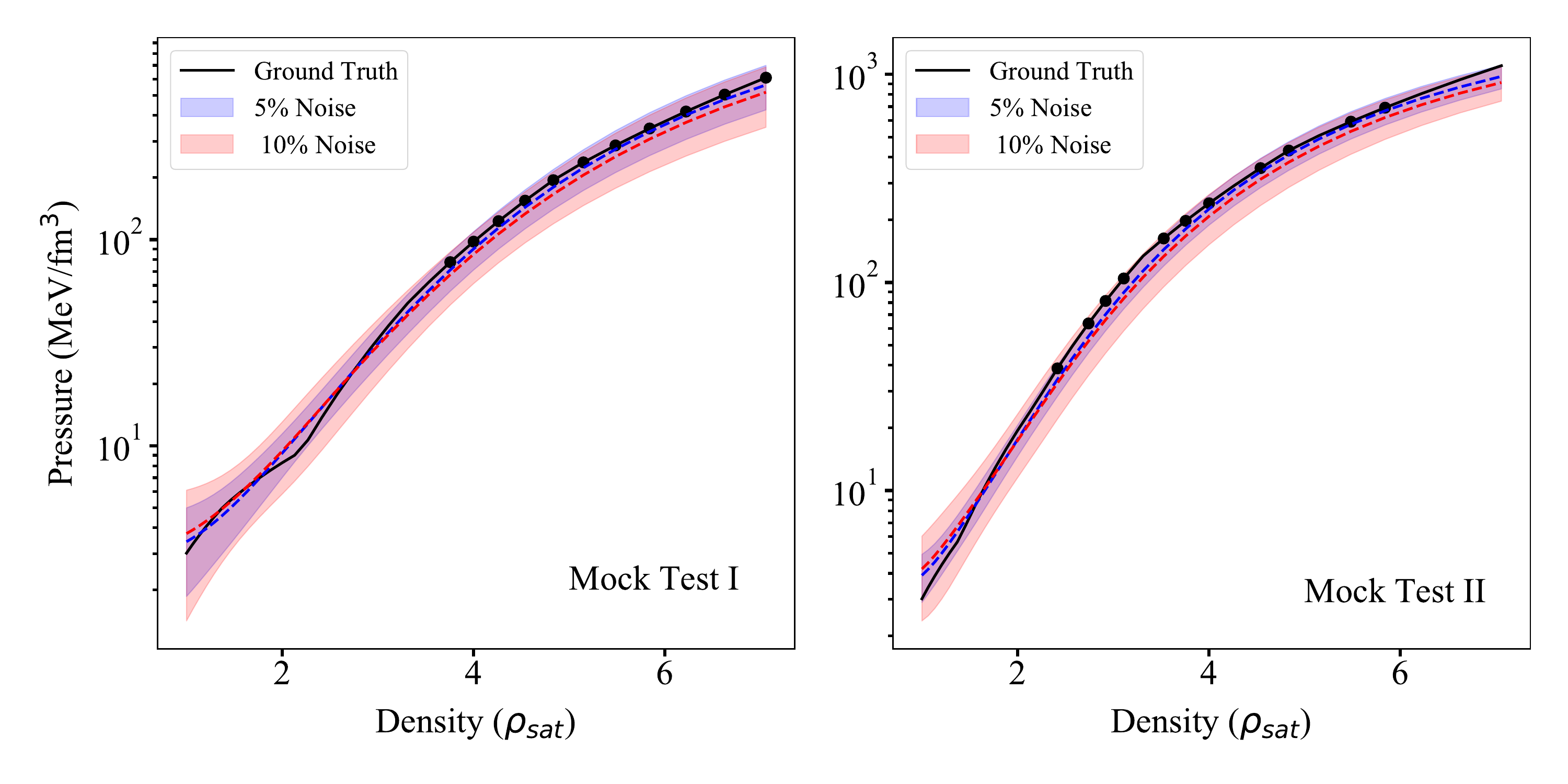}
\caption{The EoS reconstruction under different levels of uncertainties on the M-R mock observations, with the blue band denoting a $5\%$ noise level and the orange band denoting a $10\%$ noise level. The corresponding dashed lines represent the respective mean of the reconstructed EoSs. The left and right panels respectively represent the Mock Test I and Mock Test II.}\label{fig:eos_noise}
\end{center}
\end{figure}

Figure~\ref{rec_eos_mr2} and ~\ref{fig:eos_noise} show the performance of the EoS reconstruction for both mock test I and II when different observational uncertainties are assumed:
the relative noise level of mock data are set to 10\% and 5\% of the ``observed'' mean. The systematic error can, in this case, be quantified from both the uncertainty of the \texttt{TOV-Solver Network}, and of the \texttt{EoS Network}. The results shown in Figure~\ref{fig:eos_noise} demonstrate that the mean of the reconstructed EoSs is closer to the ground truth in the case with 5\% error as compared to the case with 10\% error. The uncertainty band narrows down with decreasing error. Hence, the availability of precise future observations can provide great scope for the reconstruction of a better constrained EoS using the present method.
The closure tests above demonstrate that the present AD based method with DNNs reliably reconstructs the underlying dense NS matter EoS, given limited NS M-R observations, with substantial measurement errors. This method applies an unbiased representation for the EoS with DNNs, and thus introduces an unique model-independence into the exploration of the properties of dense matter from NS structural observations. While Bayesian approaches can use polytropic EoSs for a model-independent reconstruction, the optimization in a multi-dimensional parameter space is computationally inefficient. Moreover, as the AD framework uses a pre-trained DNN to emulate the TOV equations, the method proposed here is fast and efficient. When compared to supervised learning approaches on the same problem, the present new method possesses a natural Bayesian picture for its interpretation and for estimating the uncertainties related to the observational noise, besides providing a novel alternative for the EoS reconstruction. This work can be further extended to incorporate a Bayesian Neural Network. 
An interesting aspect which can be explored is the inclusion of EoSs which undergo a first-order phase transition (FOPT) or a crossover from hadronic matter to quark matter. There is a possibility of such an occurrence of a FOPT in the cores of NSs, which may have dramatic consequences also for binary neutron star mergers, BNSMs~\cite{Most:2018eaw}, and for core collapse supernovae. In the current study we do not include such EoSs (e.g., constant pressure with increasing energy density) in the data-set.

A similar study can also be carried out by using the M-$\Lambda$ relationships of NSs. The tidal deformability estimates obtained from GWs however, have even larger uncertainties. Future measurements from next generation GW detectors, like LIGO India, KAGRA, Einstein Telescope, Cosmic Explorer and LISA, with greater sensitivities will yield better constrained estimates for tidal deformability. 
The present study is restricted to M-R observations of long-lived cold neutron stars, and tested on synthetic mock data (investigation based on real data is in progress). The study can be further extended to observations from proto-neutron stars and to remnants of BNSMs, where the high temperatures cannot be ignored. The thermal contributions in the pressure can be incorporated separately in a piecewise polytropic approach, such as, $P = P_{\text{cold}} + P_{\text{thermal}}$. The adiabatic index of the thermal pressure has been examined in earlier works and can be set to $\Gamma_{\text{th}} \sim ~$1.5-2.0~\cite{thermalgammaPhysRevD.102.043006}.
With an increase in the number of GW event detections, there is future scope for the study of long-lived remnants from BNSMs using a non-zero temperature EoS.

\section{Conclusions} 
\label{conclusions}

This work introduces a novel Automatic Differentiation (AD)-based approach with DNNs to reconstruct the EoS from stellar observations. In a Bayesian inference picture, the EoS is represented by DNNs (\texttt{EoS Network}) without explicit physical model priors. In combination with the other supervised trained DNNs (\texttt{TOV-Solver Network}), which emulate the TOV equations solver, the AD is applied to reconstruct the EoS using a gradient-based optimization by fitting NS M-R mock data. We first train the \texttt{TOV-Solver Network} and show that it can successfully replace the numerical ODE solvers, in fact, in a more efficient and easily differentiable manner. Several closure tests with mock M-R data show that the proposed approach effectively reconstructs the underlying EoS and also evaluates the associated uncertainties. 
Furthermore, we demonstrate that a higher precision on future measurements of NS global properties with next-generation telescopes and detectors can provide the scope for a fine reconstruction of the EoS of NS matter. This also proves the reliability of the proposed method. 
\\

\section{Acknowledgements}

The authors thank Dr.~Jan Steinheimer and Dr.~Anton Motornenko for useful discussions. This work is supported by (i)~Deutscher Akademischer Austauschdienst - DAAD, and GSI-F\&E funds (S. Soma), (ii)~BMBF under the ErUM-Data project (K. Zhou), (iii)~the AI grant of SAMSON AG, Frankfurt (S. Soma, L. Wang and K. Zhou), (iv)~Xidian-FIAS International Joint Research Center (L. Wang), (v)~Natural Sciences and Engineering Research Council of Canada (S. Shi), (vi)~Bourses d'excellence pour \'etudiants \'etrangers (PBEEE) from Le Fonds de Recherche du Qu\'ebec - Nature et technologies (FRQNT) (S. Shi), (vii)~U.S. Department of Energy, Office of Science, Office of Nuclear Physics, grant No. DE-FG88ER40388 (S. Shi), and (viii)~Walter Greiner Gesellschaft zur F\"orderung der physikalischen Grundlagenforschung e.V. through the Judah M. Eisenberg Laureatus Chair at Goethe Universit\"at Frankfurt am Main (H. St\"ocker). We also thank the NVIDIA Corporation for donation of NVIDIA GPUs.
\bibliographystyle{JHEP}
\bibliography{nnEoS}

\newpage
\appendix 

\section{TOV-Solver Network} \label{appendixA}

In order to train \texttt{TOV-Solver Network}, 52,000 EoS samples are taken from each set of polytropic EoSs. This accounts to 156,000 training samples corresponding to the three sets of EoSs for lower density region, which are shuffled before training. The hyperparameters of the network are initially modulated to acquire a working model. The network weights are tuned during the training process to reduce the mean squared error (MSE) between the output and the true M-R curve, which serves as the loss function for this regression task. The MSE is averaged over both the mass and radius with equal weights. Accuracy of the model is quantified by the coefficient of determination, $\mathcal{R}^2$, which is defined as $ 1 - {\sum_i(y_i - \hat{y_i})^2}/({\sum_{i}(y_i - \bar{y})^2 + \delta})$. Here, $y_i$ is the true value of the $i^{\text{th}}$ M-R pair, $\hat{y_i}$ is the corresponding prediction from the network, and $\bar{y}$ is the mean of the true M-R values. We set $\delta=10^{-7}$ to avoid undefined values when encountered with a division by zero. Once the loss converges, the weights are finalized and we are left with a well-functioning generic model. The remaining 72,569 EoS samples, exclusive of the training set, are then used for the validation and testing of the network.

Illustrated in the sections below are detailed versions of four unique types of DL models that were constructed, trained and then tested to perform this task. The input and output layers of the networks are of the same length. The dimensions, however, are different. The input layer has one channel, corresponding to the pressure $P(\rho)$, and the output layer contains two channels for the M-R pairs. Each type of model was built with two different resolutions for the data-set, i.e, $N_{\rho}=128 \text{ and } 32$; resulting in eight distinct models. We used the Adam optimizer~\cite{kingma2017adam}, with learning rate, $\alpha=0.0003$ for all the models. They were all trained by minimizing the loss function, MSE, in batches of size 4096. The accuracy of all the models was estimated by the $\mathcal{R}^2$ metric, and reaches around 99.9\% in each case. Further specifics are mentioned in their respective modules below, and a comparison of the models' functioning is done towards the end of Appendix \ref{appendixA}.

\subsection{Fully Connected Dense Neural Network (FCN)}
A fully connected dense neural network is a feed-forward network with fully connected layers, meaning that every single neuron in a layer (hidden or output) is connected to the preceding layer's neurons. The models pertaining to our task have the architectures as portrayed in Table~\ref{dnn} for  both $N_{\rho}$ = 128 and 32. In both cases, the ELU activation function was applied on the hidden layers, and Sigmoid on the output layer. The model weights were initialized with the {\it{He}} Normal distribution~\cite{he2015delving}. An $L_2$ regularization penalty ($\lambda = 10^{-7}$) was applied on the layer weights. The FCN models were trained for 15,000 epochs, given their significantly superior training times when compared to the rest of the DL models. Further details of performance are given in Table~\ref{comparison32}.
\begin{table}[ht!]
\caption{The FCN model architecture used for the \texttt{TOV-Solver Network}.}
\centering
\def\arraystretch{1.2}
\setlength\tabcolsep{10pt}
\begin{tabular}{@{}cccc@{}}
\hline\hline
Layer Index & Layer & \multicolumn{2}{c}{Dimension} \\
\hline
            &       & $N_{\rho}$ = 128 & $N_{\rho}$ = 32 \\
\hline%
Input & -    & 128 & 32 \\
1 & Dense    & 128 & 32 \\
2 & Dense    & 64  & 64 \\ 
3 & Dense    & 32  & 128 \\ 
4 & Dense    & 64  & 64 \\ 
- & Add (2,4)& 64  & 64 \\
5 & Dense    & 128 & 32 \\ 
- & Add (1,5)& 128 & 32 \\
6 & Dense    & 256 & 64 \\
Output & Reshape & (128,2) & (32,2) \\
\hline\hline
\end{tabular}
\label{dnn}
\end{table}
\subsection{Convolutional Neural Network (CNN)}
CNNs are widely used as they have the potential to extract key features or patterns in the input data. This type of a DL model is explored to make certain the network is not rendered obsolete when encountered with stark changes in the pressure gradient. The CNN consists of a convolutional kernal or matrix for each layer, which is used to compute the outputs of the succeeding layer~\cite{Nielsen2015}. The number of layers in the two  CNN models that we built and their dimensions are listed in Table~\ref{cnn}. A kernel size of 3$\times$1 is used for both the models, with the weights initialized from the {\it{He}} Normal distribution~\cite{he2015delving}, and penalized with the $L_2$ regularizer ($\lambda = 10^{-6}$). Stride values of 1 and 2 are applied alternatively on the convolutional layers. We employ the ELU activation function on the hidden layers and Sigmoid on the output layer. The padding {\it `same'} is applied to the convolutional layers to disallow loss of information at the boundaries or a change in dimension of the following layer. 
\begin{table}[ht!]
\caption{The CNN model architecture used for the \texttt{TOV-Solver Network}.}
\centering
\def\arraystretch{1.2}
\setlength\tabcolsep{10pt}
\begin{tabular}{@{}cccc@{}}
\hline\hline
Layer Index & Layer & \multicolumn{2}{c}{Dimension} \\
\hline
            &       & $N_{\rho}$ = 128 & $N_{\rho}$ = 32 \\
\hline%
Input & - & (128,1) & (32,1)\\
1 & Convolution 1D & (128,128) & (32,32) \\
2 & Convolution 1D & (64,64)  & (16,64)\\ 
3 & Convolution 1D & (64,64)  & (16,64)\\ 
- & Add (2,3) & (64,64)       & (16,64)\\
4 & Convolution 1D & (32,64)  & (8,64)\\ 
5 & Convolution 1D & (32,64)  & (8,64)\\ 
- & Add (4,5) & (32,64)       & (8,64)\\
6 & Convolution 1D & (16,32)  & (4,32)\\
7 & Convolution 1D & (16,32)  & (4,32)\\
- & Add (6,7) & (16,32)      & (4,32)\\
- & Reshape & 512  & 128\\
8 & Dense & 128 & 32\\
9 & Dense & 256 & 64\\
Output & Reshape & (128,2) & (32,2)\\
\hline\hline
\end{tabular}
\label{cnn}
\end{table}
\subsection{Long Short Term Memory (LSTM) Network}
The Long Short Term Memory, or LSTM, is a recurrent NN, which as the name suggests, has recurrent connections or feedback loops in a layer, in order to retain memory in a sequence~\cite{10.1162/neco.1997.9.8.1735}. The recurring units or memory cells in LSTM carry the dependency across time sequences, leading to longer training times (see Table~\ref{comparison32} for a comparison to other models). Due to the fact that the output layer is an M-R sequence, the prospects of LSTM for the \texttt{TOV-Solver Network} are examined. The model descriptions are given in Table~\ref{lstm} for $N_{\rho}$ = 128, 32. The kernel parameters of all the layers were initialized from the Xavier or Glorot uniform distribution~\cite{glorot2010understanding}, and the $L_2$ regularization ($\lambda = 10^{-7}$) was applied. In this case, we used the Tanh activation function for the LSTM layers.

\begin{table}[ht!]
\caption{The LSTM model architecture used for the \texttt{TOV-Solver Network}.}
\centering
\def\arraystretch{1.2}
\setlength\tabcolsep{10pt}
\begin{tabular}{@{}cccc@{}}
\hline\hline
Layer Index & Layer & \multicolumn{2}{c}{Dimension} \\
\hline
            &       & $N_{\rho}$ = 128 & $N_{\rho}$ = 32 \\
\hline%
Input & - & (128,1)  & (32,1)\\
1 & LSTM & (128,128) & (32,32)\\
2 & LSTM & (128,64) & (32,64) \\ 
3 & LSTM & (128,128) & (32,64)\\ 
4 & LSTM & (128,128) & (32,32)\\ 
- & Add (1,4) & (128,128) & (32,32)\\
5 (Output) & LSTM & (128,2) & (32,2)\\ 
\hline\hline
\end{tabular}
\label{lstm}
\end{table}
\subsection{WaveNet}
The WaveNet is a generative NN model with autoregressive properties. The network is structured with convolutional layers that make use of dilated kernels~\cite{10.1007/978-3-642-75988-8_28}. 
The hidden layers with inflated kernels along with {\it `causal'} padding ensure the autoregressive behaviour of the network, without forfeiting any information. The network mimics the concept of autoregression that is exercised in solving the TOV equations, and is therefore used in further stages of our research study. The kernel parameters are initialized from the Xavier or Glorot uniform distribution~\cite{glorot2010understanding}, and the $L_2$ regularization ($\lambda = 10^{-7}$) is applied. The ELU activation function is applied on the all the layers but the last (Sigmoid activation). 

\begin{table}[ht!]
\caption{The WaveNet model architecture used for the \texttt{TOV-Solver Network}.}
\centering
\def\arraystretch{1.2}
\setlength\tabcolsep{6pt}

\begin{tabular}{@{}ccccc@{}}
\hline\hline
Layer Index & Layer & Dilation & \multicolumn{2}{c}{Dimension} \\
\hline
            &      &       & $N_{\rho}$ = 128 & $N_{\rho}$ = 32 \\
\hline%
Input & - & - & (128,1) & (32,1)\\
1 & Convolution 1D & - & (128,128) & (32,32)\\
2 & Convolution 1D & 1 & (128,128) & (32,32)\\ 
3 & Convolution 1D & 2 & (128,128) & (32,32)\\ 
4 & Convolution 1D & 4 & (128,128) & (32,32)\\ 
5 & Convolution 1D & 8 & (128,128) & (32,32)\\ 
6 & Convolution 1D & 16 & (128,128) & (32,32)\\
7 & Convolution 1D & 32 & (128,128) & (32,32)\\
8 & Convolution 1D & 16 & (128,128) & (32,32)\\
9 & Convolution 1D & 32 & (128,128) & (32,32)\\
10 (Output) & Convolution 1D & 64 & (128,2) & (32,2)\\
\hline\hline
\end{tabular}
\label{wn}
\end{table}

\subsection{Comparison}
The performance of the different models was examined on a validation set, which is exclusive of the training data. The tables show that given EoSs, all the DNNs are capable of finding their TOV solutions at precisions that reach $\mathcal{R}^2$=99.9\%.
The LSTM and WaveNet models are typically used for training series or sequences. Therefore, we observe a boost in their performances when combined with the \texttt{EoS-Network}. 

\begin{table}[ht]
\caption{Comparison of the performance of different Neural Networks for solving the TOV equations.} 
\centering
\def\arraystretch{1.4}
\setlength\tabcolsep{2pt}
\begin{tabular}{c| c c c c c c c c c c}
\hline\hline
$N_{\rho}$ & NN & $\mathcal{R}^2$ & MSE & Parameters & Epochs  & Time \\
& & &  $(\times 10^{-5})$ & ($\#$) & ($\times 10^3$)  &($\times 10^3 sec$)\\
 \hline
\multirow{4}{*}{128}& CNN     &0.9999 &1.743 &170,176 &3.5  &7.35  \\ 
& FCN     &0.9999 &1.052 &70,304  &15   &4.91  \\
& LSTM    &0.9998 &0.741 &347,416 &3    &32.5 \\  
& WaveNet &0.9998 &3.003 &296,706 &3    &64.7 \\  
 \hline
\multirow{4}{*}{32} & CNN     &0.9999 &3.019 &58,912 &3.5 &2.15   \\ 
 & FCN     &0.9999 &1.179 &23,936 &15  &2.79   \\  
 & LSTM    &0.9999 &0.814 &74,904 & 3  &4.03   \\  
 & WaveNet &0.9999 &3.047 &18,882 & 3  &10.7  \\  
\hline\hline
\end{tabular}
\label{comparison32}
\end{table}

\section{Mock Tests on SFHo and BHB$\Lambda\phi$ EoSs.} \label{AppendixB}
\begin{figure}[tbp]
    \centering
    \includegraphics[width=0.45\textwidth]{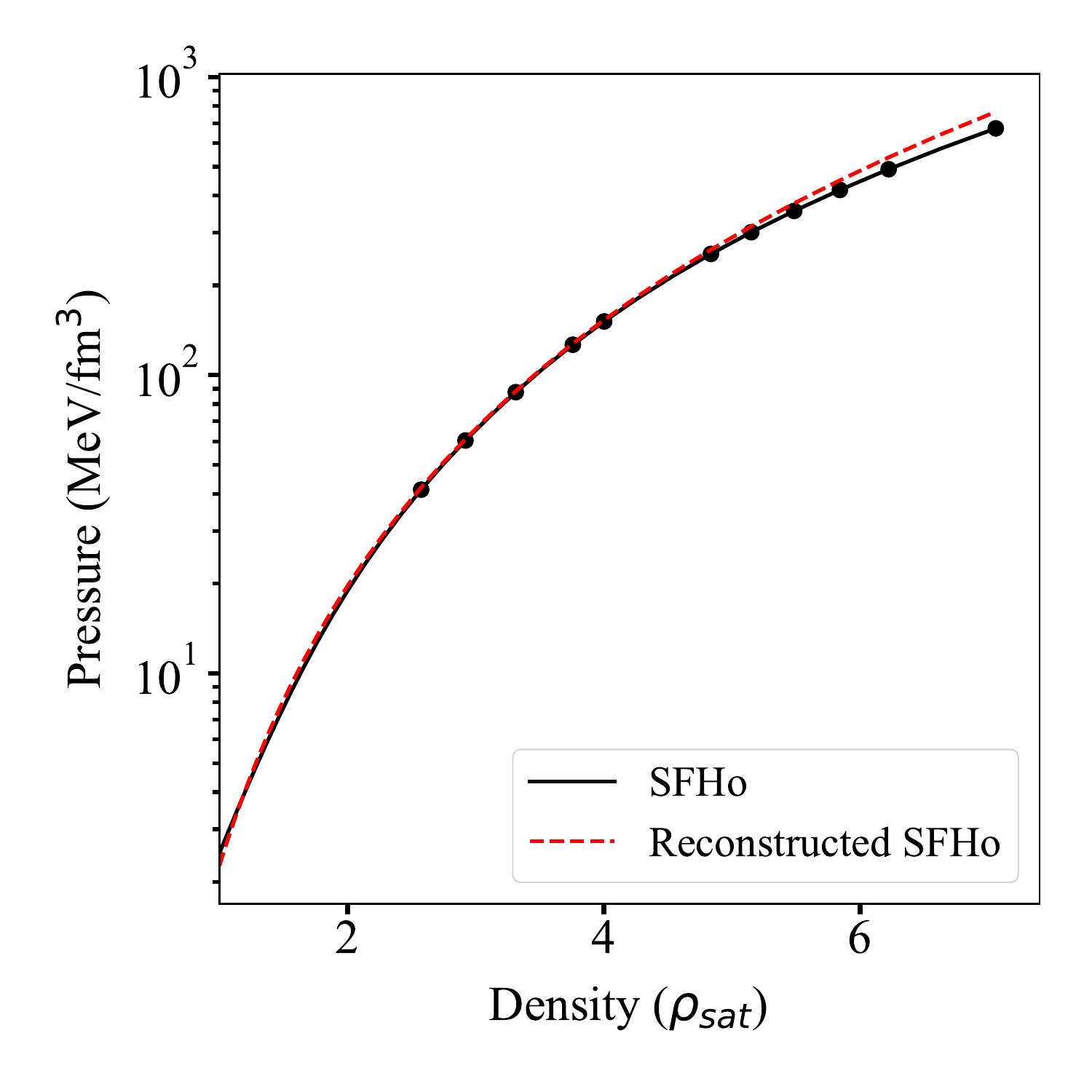}
    \hfill    
    \includegraphics[width=0.45\textwidth]{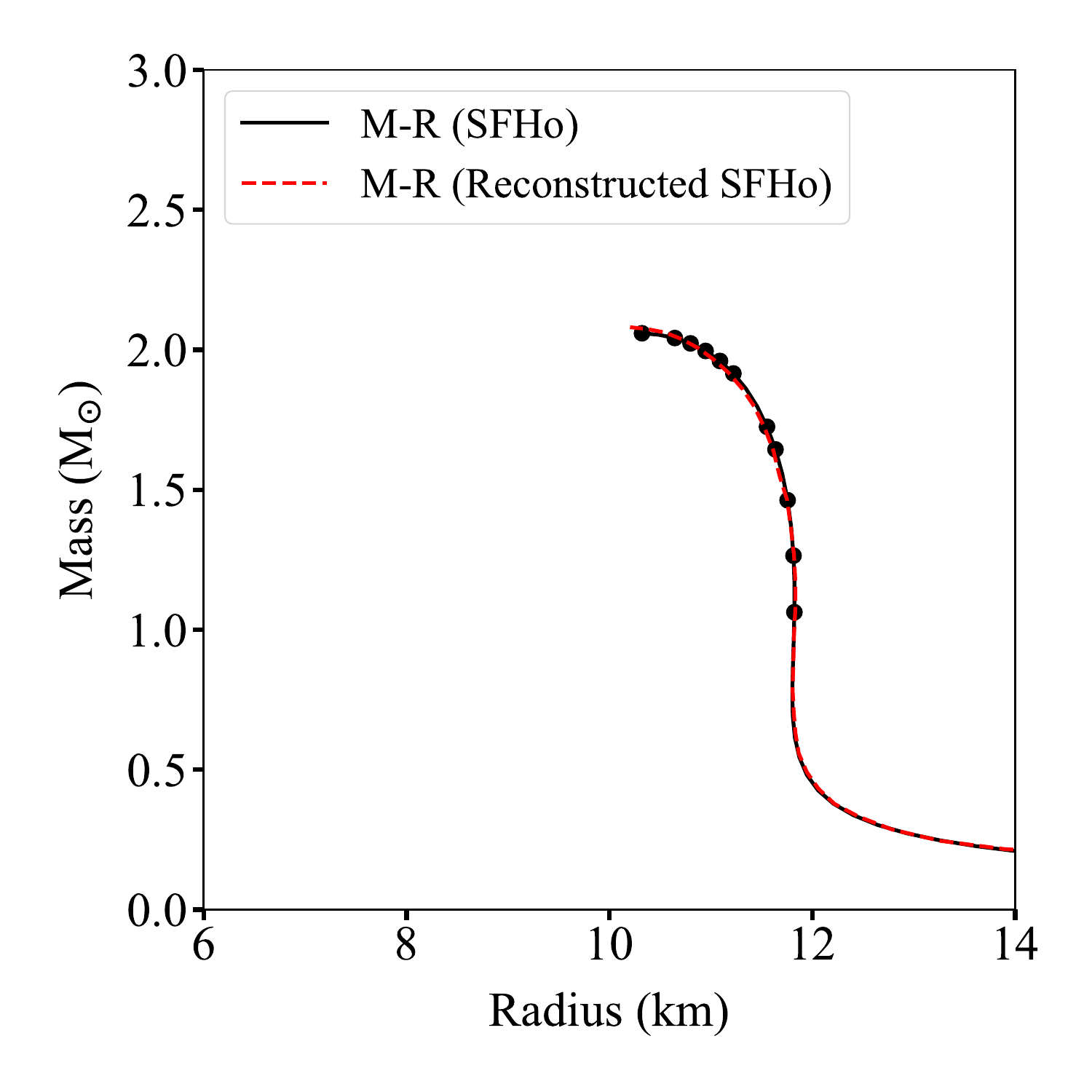}
    \caption{Reconstruction performance check on the SFHo EoS.}
    \label{fig:reconsfho}
\end{figure}

\begin{figure}[tbp]
    \centering
    \includegraphics[width=0.45\textwidth]{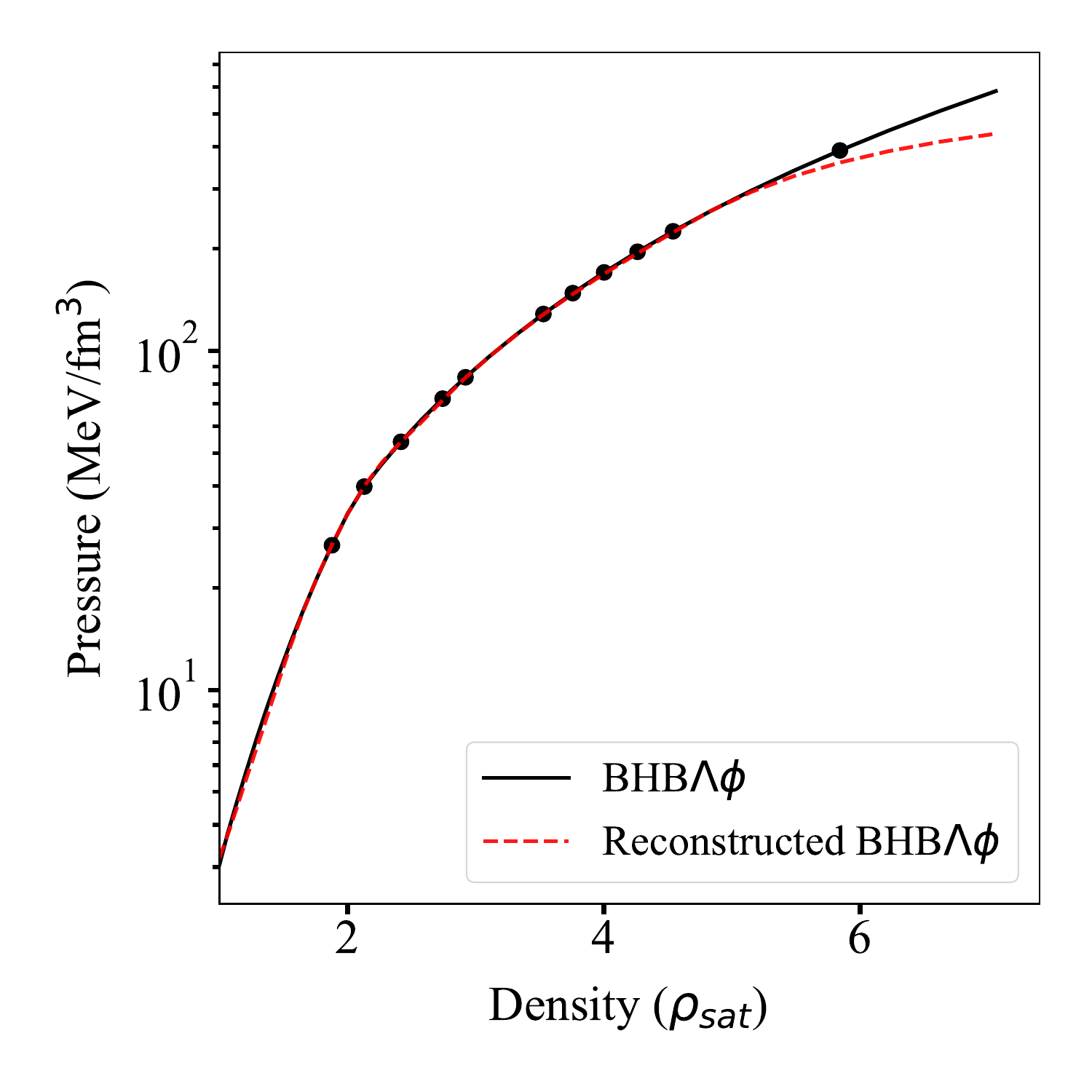}
    \hfill    
    \includegraphics[width=0.45\textwidth]{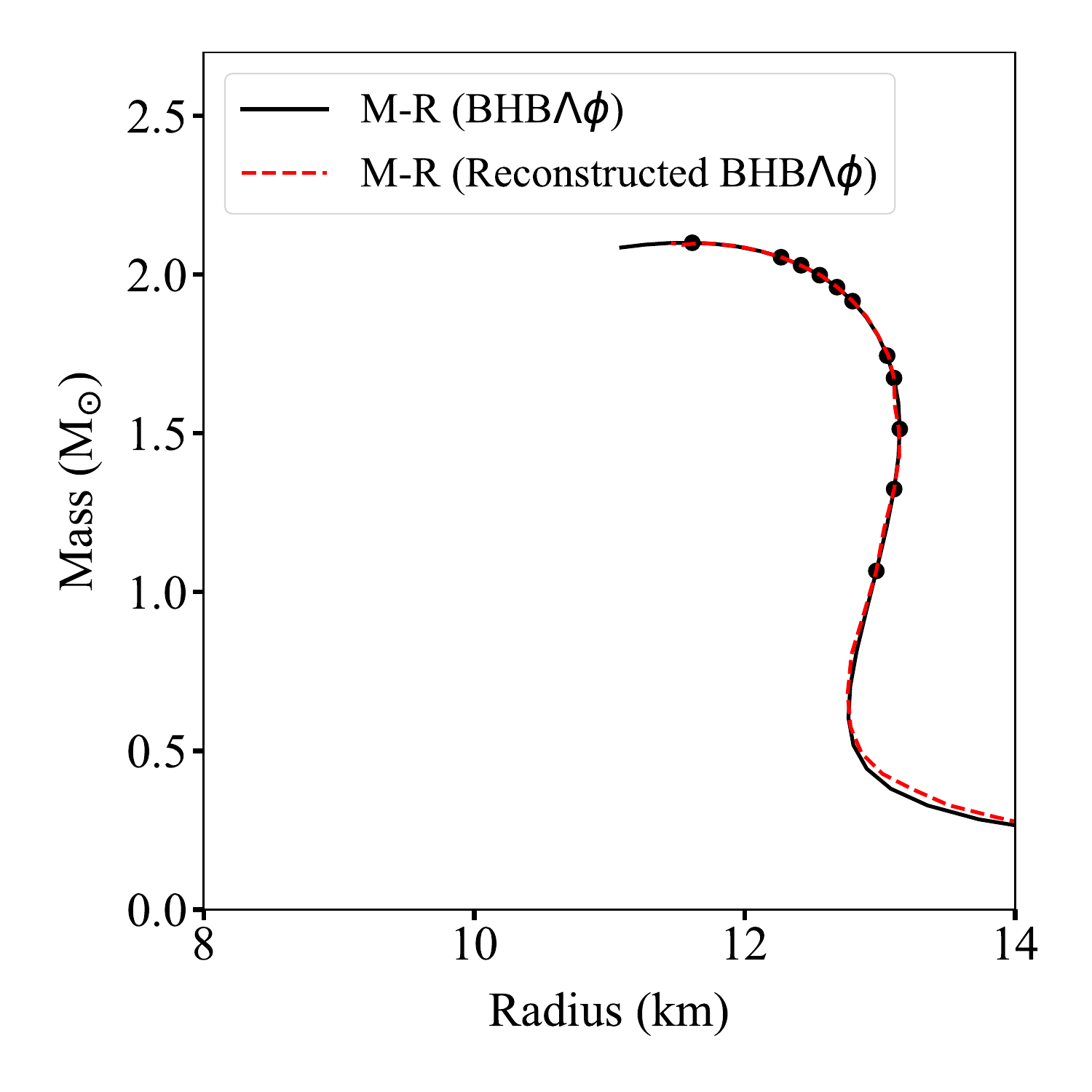}
    \caption{Reconstruction performance check on the BHB$\Lambda\phi$ EoS.}
    \label{fig:reconbhblp}
\end{figure}

Here, we show the performance of the neural networks on the reconstruction of SFHo and BHB$\Lambda\phi$ EoSs, using the proposed method. The mock M-R observational data points are taken from corresponding M-R curve obtained with the SFHo or BHB$\Lambda\phi$ EoSs. We perform the tests without assuming uncertainties. The reconstructed SFHo and BHB$\Lambda\phi$ EoSs obtained from the optimization procedure are depicted as red dashed curves in the left panels of Figure \ref{fig:reconsfho} and Figure \ref{fig:reconbhblp} respectively. The results are compared with the true EoSs which are depicted as solid black curves. The M-R curves obtained from the reconstructed EoSs (red dashed curves) are also shown against the true M-R curves (solid black curves) in the right panels of Figure \ref{fig:reconsfho} and Figure \ref{fig:reconbhblp}.


\end{document}